\documentclass[12pt]{article}
\usepackage{enumerate}
\usepackage{amsfonts}
\usepackage{amsmath}
\usepackage{amssymb}
\usepackage{amsthm}
\usepackage{latexsym}
\usepackage{color}

\def\H{\mathcal{H}}

\def\M{\mathbb{M}}

\def\S{\mathfrak{S}}

\def\F{\mathfrak{F}}
\def\T{\mathfrak{T}}
\def\B{\mathfrak{B}}

\newcommand{\rank}{\mathrm{rank}}
\newcommand{\id}{\mathrm{Id}}
\newcommand{\Tr}{\mathrm{Tr}}

\newcommand{\shs}{\hspace{1pt}}
\newcounter{defin}  \newcounter{lemma}  \newcounter{theorem}
\newcounter{property} \newcounter{corol}  \newcounter{remark} \newcounter{example}
\newenvironment{lemma}{\par\refstepcounter{lemma}     \textbf{Lemma \thelemma.} }{\rm\par}
\newenvironment{theorem}{\par\refstepcounter{theorem}     \textbf{Theorem \thetheorem.}\ }{\rm\par}
\newenvironment{property}{\par\refstepcounter{property}     \textbf{Proposition \theproperty.}\ }{\rm\par}
\newenvironment{corollary}{\par\refstepcounter{corol}     \textbf{Corollary \thecorol.} }{\rm\par}

\newenvironment{example}{\par\refstepcounter{example}     \textbf{Example \theexample.}}{\rm\par}

\begin{document}

\title{On lower semicontinuity of the quantum conditional mutual information and its corollaries}

\author{M.E. Shirokov\footnote{Steklov Mathematical Institute, RAS, Moscow, email:msh@mi.ras.ru}}
\date{}
\maketitle

\begin{abstract}
It is well known that the quantum mutual information and its conditional version do not increase under  local channels. I this paper we show
that the recently established lower semicontinuity of the quantum conditional mutual information implies (in fact, is equivalent to) the
lower semicontinuity of the loss of the quantum (conditional) mutual information under  local channels considered as a function on the Cartesian product of the set of all states of a composite system and the sets of all local channels (equipped with the strong convergence).\smallskip

Some applications of this property are considered. New continuity conditions for the quantum mutual information and for the squashed entanglement in both bipartite and multipartite infinite-dimensional systems are obtained. It is proved, in particular, that the multipartite squashed entanglement of any countably-non-decomposable separable state with finite marginal entropies is equal to zero.

Special continuity properties of the information gain of a quantum measurement with and without quantum side information are established that can be treated as robustness (stability) of these quantities w.r.t. perturbation of the measurement and the measured state.
\end{abstract}

\tableofcontents

\section{Introduction}

The quantum mutual information and its conditional version are basic entropic quantities essentially used in different tasks of quantum information theory \cite{H-SCI,N&Ch,Wilde}. In finite dimensions these quantities are continuous bounded functions on the set of all states of a composite quantum system. In the case of bipartite system $AB$ and conditioning system $C$ they are defined by the expressions
\begin{equation}\label{mi-1}
I(A\!:\!B)_{\rho}=H(\rho_{A})+H(\rho_{B})-H(\rho_{AB})
\end{equation}
and
\begin{equation}\label{cmi-1}
I(A\!:\!B|C)_{\rho}=H(\rho_{AC})+H(\rho_{BC})-H(\rho_{ABC})-H(\rho_{C}).
\end{equation}
In the infinite-dimensional case these formulae may contain the uncertainty $+\infty-\infty$. So, in this case one should use
expressions for $I(A\!:\!B)_{\rho}$ and $I(A\!:\!B|C)_{\rho}$ which are less sensitive to infinite values of marginal entropies.

It is well known that the quantum mutual information has a nonnegative lower semicontinuous extension to the set of all states of infinite-dimensional
composite system. In the bipartite case this extension is given by the formula
$$
I(A\!:\!B)_{\rho}=H(\rho_{AB}\shs\Vert\shs\rho_{A}\otimes
\rho_{\shs B}),
$$
where $H(\cdot\Vert\cdot)$ is the quantum relative entropy, coinciding with (\ref{mi-1}) for a state $\rho_{AB}$ with finite marginal entropies $H(\rho_{A})$ and $H(\rho_{B})$ \cite{L-mi}.

For the quantum conditional mutual information (QCMI) there are several expressions permitting to define it for states with infinite marginal entropies. For example, in the tripartite case we have
$$
I(A\!:\!B|C)_{\rho}=I(AC\!:\!B)_{\rho}-I(B\!:\!C)_{\rho}\quad \textrm{and} \quad I(A\!:\!B|C)_{\rho}=I(A\!:\!BC)_{\rho}-I(A\!:\!C)_{\rho}.
$$
But these and similar expressions (see \cite[Section 6]{CMI}) are not well defined on whole space of states of an infinite-dimensional composite system.

Nevertheless, it is shown in \cite{CMI} that in both tripartite and multipartite cases there exists a unique lower semicontinuous function well defined on the whole space of states of an infinite-dimensional composite system possessing all basic properties of QCMI (in particular, coinciding with the
original definition (\ref{cmi-1}) and all alternative expressions for QCMI at states, where they are well defined). Unfortunately, this function can not be
expressed by a simple formula, it is defined via some optimization procedure (see details in Section 3).

Despite the absent of explicit formula, the existence of a unique extension of QCMI to the whole
space of states of an infinite-dimensional composite system is important because it allows to not care about the characteristics of the state at  which
QCMI is calculated in formal manipulations. For example, when we apply a quantum channel to a part of a composite state at which QCMI is well defined by a
simple formula we may obtain a state at which this formula does not work, but the existence of global extension of QCMI allows to forget about this problem. By this reason the global extension of QCMI plays a central role in the resent works \cite{CID,UFA}, where the main results are formulated as
a property valid for any channel between given quantum systems: the uniform continuity of basic capacities on the set of all channels, the uniform finite-dimensional approximations of basic capacities, etc. The global extension of QCMI is also used essentially in the recent work  \cite{D-S-W}.

In this paper we focus attention on the \emph{lower semicontinuity} of the global extension of QCMI mentioned before on the whole
space of states of an infinite-dimensional composite system. In the tripartite case it means that
$$
\liminf_{n\to +\infty}I(A\!:\!B|C)_{\rho^n}\geq I(A\!:\!B|C)_{\rho^0}
$$
for any sequence of states $\rho^n_{ABC}$ converging to a state $\rho^0_{ABC}$. It is equivalent to the closedness  of the set of
all states $\rho_{ABC}$ determined by the inequality
$$
I(A\!:\!B|C)_{\rho}\leq \epsilon
$$
for any nonnegative $\epsilon$. For $\epsilon=0$ this means the closedness of the set of short Markov chains \cite{H&Co+}, which by no means is obvious in infinite dimensions.

By using the lower semicontinuity of QCMI and the Stinespring representation of strongly converging sequences of quantum channels
obtained in \cite{CSR} we prove (in Section 4) the lower semicontinuity of the loss of the quantum (conditional) mutual information under action of local channels considered as a function on the Cartesian product of the set of all states of a composite system and the sets of all local channels (equipped with the strong convergence topology). In the tripartite case it means that the function\vspace{-10pt}
$$
(\rho,\Phi_A,\Phi_B)\mapsto I(A\!:\!B|C)_{\rho}-I(A'\!:\!B'|C)_{\Phi_A\otimes\Phi_B\otimes\id_{\!C}(\rho)},
$$
where $\Phi_X$ denotes a channel from $X$  to $X'$, is lower semicontinuous on the set of all triplets $(\rho_{ABC},\Phi_A,\Phi_B)$ such that
$I(A'\!:\!B'|C)_{\Phi_A\otimes\Phi_B\otimes\id_{\!C}(\rho)}<+\infty$ w.r.t. the convergence:
$$
\{(\rho^n_{ABC},\Phi^n_A,\Phi^n_B)\to(\rho^0_{ABC},\Phi^0_A,\Phi^0_B)\}\Leftrightarrow \{\rho^n_{ABC}\to\rho^0_{ABC}\}\wedge\{\Phi^n_A\to\Phi^0_A\}\wedge\{\Phi^n_B\to\Phi^0_B\},
$$
where $\,\Phi^n_X\to\Phi^0_X$ denotes the strong convergence of the sequence $\{\Phi^n_X\}$ to the channel $\Phi^0_X$
which  means that $\{\Phi^n_X(\varrho_X)\}$ tends to the state $\Phi^0_X(\varrho_X)$ for any state $\varrho_X$.\smallskip

The above property has a direct analogy  to the lower semicontinuity of the entropic disturbance\footnote{The entropic disturbance is the difference between the Holevo quantities of an ensemble of quantum states and the image of this ensemble under a quantum channel \cite{ED-1,ED-2}.} as function of a pair
(channel, input ensemble) established in \cite{CSR}. In fact, if we restrict attention to discrete (finite or countable) ensembles
of quantum states then the lower semicontinuity of the entropic disturbance can be derived from the lower semicontinuity of the loss
of the quantum mutual information under action of a local channel.

The lower semicontinuity of the loss
of QCMI under local channels allows to show that the action of arbitrary local channels
preserves continuity of QCMI. In the tripartite case it means that the limit relation
$$
\,\lim_{n\to+\infty}I(A\!:\!B|C)_{\rho^n}=I(A\!:\!B|C)_{\rho^0}<+\infty
$$
for some sequence of states $\rho^n_{ABC}$ converging to a state $\rho^0_{ABC}$ implies that
$$
\lim_{n\to+\infty}I(A'\!:\!B'|C)_{\Phi_A\otimes\Phi_B\otimes\id_{\!C}(\rho^n)}=I(A'\!:\!B'|C)_{\Phi_A\otimes\Phi_B\otimes\id_{\!C}(\rho^0)}
$$
for \emph{arbitrary} channels $\Phi_A:A\rightarrow A'$ and $\Phi_B:B\rightarrow B'$. Moreover, this result can be strengthened by replacing
fixed channels $\Phi_A$ and $\Phi_B$ by strongly converging sequences of channels (see Proposition \ref{main-r} in Section 5.1.1)
\smallskip

In Section 5 we show how the above "preserving continuity" property can be used in continuity analysis of QCMI and relating important characteristics
of quantum states and quantum measurements.

\section{Preliminaries}

Let $\mathcal{H}$ be a separable Hilbert space,
$\mathfrak{B}(\mathcal{H})$ the algebra of all bounded operators on $\mathcal{H}$ with the operator norm $\|\cdot\|$ and $\mathfrak{T}( \mathcal{H})$ the
Banach space of all trace-class
operators on $\mathcal{H}$  with the trace norm $\|\!\cdot\!\|_1$. Let
$\mathfrak{S}(\mathcal{H})$ be  the set of quantum states (positive operators
in $\mathfrak{T}(\mathcal{H})$ with unit trace) \cite{H-SCI,N&Ch,Wilde}.

Denote by $I_{\mathcal{H}}$ the unit operator on a Hilbert space
$\mathcal{H}$ and by $\id_{\mathcal{\H}}$ the identity
transformation of the Banach space $\mathfrak{T}(\mathcal{H})$.\smallskip

The \emph{von Neumann entropy} of a quantum state
$\rho \in \mathfrak{S}(\H)$ is  defined by the formula
$H(\rho)=\operatorname{Tr}\eta(\rho)$, where  $\eta(x)=-x\log x$ for $x>0$
and $\eta(0)=0$. It is a concave lower semicontinuous function on the set~$\mathfrak{S}(\H)$ taking values in~$[0,+\infty]$ \cite{H-SCI,L-2,W}.
The von Neumann entropy satisfies the inequality
\begin{equation}\label{w-k-ineq}
H(p\rho+(1-p)\sigma)\leq pH(\rho)+(1-p)H(\sigma)+h_2(p)
\end{equation}
valid for any states  $\rho$ and $\sigma$ in $\S(\H)$ and $p\in(0,1)$, where $\,h_2(p)=\eta(p)+\eta(1-p)\,$ is the binary entropy \cite{N&Ch,Wilde}.\smallskip

The \emph{quantum relative entropy} for two states $\rho$ and
$\sigma$ in $\mathfrak{S}(\mathcal{H})$ is defined as
$$
H(\rho\,\|\shs\sigma)=\sum\langle
i|\,\rho\log\rho-\rho\log\sigma\,|i\rangle,
$$
where $\{|i\rangle\}$ is the orthonormal basis of
eigenvectors of the state $\rho$ and it is assumed that
$H(\rho\,\|\sigma)=+\infty$ if $\,\mathrm{supp}\rho\shs$ is not
contained in $\shs\mathrm{supp}\shs\sigma$ \cite{H-SCI,L-2}.\footnote{The support $\mathrm{supp}\rho$ of a state $\rho$ is the closed subspace spanned by the eigenvectors of $\rho$ corresponding to its positive eigenvalues.}\smallskip

The \emph{quantum conditional entropy}
\begin{equation}\label{c-e-d}
H(A|B)_{\rho}=H(\rho_{AB})-H(\rho_{\shs B})
\end{equation}
of a  state $\rho_{AB}$ with finite marginal entropies is essentially used in analysis of quantum systems \cite{H-SCI,Wilde}. It
can be extended to the set of all states $\rho_{AB}$ with finite $H(\rho_A)$ by the formula
\begin{equation}\label{ce-ext}
H(A|B)=H(\rho_{A})-H(\rho_{AB}\shs\Vert\shs\rho_{A}\otimes
\rho_{B}).
\end{equation}
This extension  possesses all basic properties of the quantum conditional entropy valid in finite dimensions \cite{Kuz}.\smallskip

The \emph{quantum mutual information} of a state $\,\rho_{AB}\,$ is defined as
\begin{equation}\label{mi-d}
I(A\!:\!B)_{\rho}=H(\rho_{AB}\shs\Vert\shs\rho_{A}\otimes
\rho_{\shs B})=H(\rho_{A})+H(\rho_{\shs B})-H(\rho_{AB}),
\end{equation}
where the second formula is valid if $\,H(\rho_{AB})\,$ is finite \cite{L-mi}.
Basic properties of the relative entropy show that $\,\rho_{AB}\mapsto
I(A\!:\!B)_{\rho}\,$ is a lower semicontinuous function on the set
$\S(\H_{AB})$ taking values in $[0,+\infty]$. It is well known that
\begin{equation}\label{MI-UB}
I(A\!:\!B)_{\rho}\leq 2\min\left\{H(\rho_A),H(\rho_B)\right\}
\end{equation}
for any state $\rho_{AB}$ \cite{L-mi,Wilde}. Local continuity of one of these upper bounds
implies local continuity of $I(A\!:\!B)$:
\begin{equation}\label{mi-cont}
\lim_{n\rightarrow\infty}H(\rho^n_X)=H(\rho^0_X)<+\infty\quad \Rightarrow \quad\lim_{n\rightarrow+\infty}I(A\!:\!B)_{\rho^n}=I(A\!:\!B)_{\rho^0}
\end{equation}
for any sequence $\{\rho^n_{AB}\}$  converging to a state $\rho^0_{AB}$, where  $X$ is either $A$ or $B$ \cite[Th.1]{CMI}.

\smallskip

A \emph{quantum channel} $\,\Phi$ from a system $A$ to a system
$B$ is a completely positive trace preserving linear map from
$\mathfrak{T}(\mathcal{H}_A)$ into $\mathfrak{T}(\mathcal{H}_B)$ \cite{H-SCI,Wilde}. For any  quantum channel $\,\Phi:A\rightarrow B\,$ the Stinespring theorem implies existence of a Hilbert space
$\mathcal{H}_E$ and of an isometry
$V_{\Phi}:\mathcal{H}_A\rightarrow\mathcal{H}_B\otimes\mathcal{H}_E$ such
that
\begin{equation*}
\Phi(\rho)=\mathrm{Tr}_{E}V_{\Phi}\rho V_{\Phi}^{*},\quad
\rho\in\mathfrak{T}(\mathcal{H}_A).
\end{equation*}
The space $\H_E$ is called \emph{environment}, its minimal dimension is called \emph{Choi rank} of the channel $\Phi$ \cite{H-SCI,Wilde}.\smallskip

Denote by $\F(A,B)$ the set of all channels from $A$ to $B$ equipped with the \emph{topology of strong convergence}  defined by the family of seminorms $\Phi\mapsto\|\Phi(\rho)\|_1$, $\rho\in\S(\H_A)$ \cite{AQC}. The strong convergence of a sequence $\{\Phi_n\}$ of channels in $\F(A,B)$ to a channel $\Phi_0\in\F(A,B)$  means that
\begin{equation*}
\lim_{n\rightarrow\infty}\Phi_n(\rho)=\Phi_0(\rho)\,\textup{ for all }\rho\in\S(\H_A).
\end{equation*}

We will use the following important result. \smallskip

\begin{lemma}\label{D-A} \cite{D-A} \emph{If a sequence $\{\rho_n\}$ of states converges to a state $\rho_0$ w.r.t. the weak operator topology then
the sequence $\{\rho_n\}$ of states converges to the state $\rho_0$ w.r.t. the trace norm.}
\end{lemma}\smallskip

We will also use the following simple\smallskip

\begin{lemma}\label{tl} \emph{If $\,h=f+g$, where $f$ and $g$ are lower semicontinuous lower bounded functions on a metric space $X$, then
continuity of $\,h$ on a subset $X_0\subseteq X$ implies continuity of $f$ and $g$ on $X_0$.}
\end{lemma}\smallskip

\section{Extended QCMI and its properties}

\subsection{Tripartite system}

The quantum conditional mutual information (QCMI) of a state $\rho_{ABC}$ of a
tripartite finite-dimensional system $ABC$ is defined as follows
\begin{equation}\label{cmi-d}
    I(A\!:\!B|C)_{\rho}\doteq
    H(\rho_{AC})+H(\rho_{BC})-H(\rho_{ABC})-H(\rho_{C}).
\end{equation}
This quantity plays important role in different areas of quantum
information theory \cite{ED-2, C&W, D&J, F&R, H&Co+, UFA, IGSI, Tucci, Wilde, Z}, it has the
following basic properties:
\begin{enumerate}[1)]
    \item $I(A\!:\!B|C)_{\rho}\geq0$ for any state $\rho_{ABC}$ and $I(A\!:\!B|C)_{\rho}=0$ if and only if there is a channel $\Phi:C\rightarrow BC$
    such that $\rho_{ABC}=\id_A\otimes\Phi(\rho_{AC})$ \cite{H&Co+};
    \item $I(A\!:\!B|C)_{\rho}=I(AC\!:\!B)_{\rho}-I(C\!:\!B)_{\rho}=I(A\!:\!BC)_{\rho}-I(A\!:\!C)_{\rho}$ for any state $\rho_{ABC}$;
    \item monotonicity under local channels: $I(A\!:\!B|C)_{\rho}\geq
    I(A'\!:\!B'|C)_{\Phi_A\otimes\Phi_B\otimes\id_{\!C}(\rho)}$ for arbitrary
    quantum channels $\Phi_A:A\rightarrow A'$ and $\Phi_B:B\rightarrow
    B'$;\footnote{The same inequality holds for any quantum operations $\Phi_A:A\rightarrow A'$ and $\Phi_B:B\rightarrow
    B'$ provided that the QCMI at any positive operator $\sigma$ is defined as $I(A\!:\!B|C)_{\sigma}=cI(A\!:\!B|C)_{\sigma/c}$, where $c=\Tr\sigma$. This can be derived from the monotonicity of relative entropy, f.i., by using formula (9.6) in \cite{CMI}.}
    \item additivity: $I(AA'\!:\!BB'|CC')_{\rho\otimes\rho'}=
    I(A\!:\!B|C)_{\rho}+I(A'\!:\!B'|C')_{\rho'}$ for any states $\rho_{ABC}$ and $\rho'_{A'B'C'}$;
    \item duality: $I(A\!:\!B|C)_{\rho}=
    I(A\!:\!B|D)_{\rho}$ for any pure state $\rho_{ABCD}$ \cite{D&J}.
\end{enumerate}\smallskip
The nonnegativity of $I(A\!:\!B|C)_{\rho}$ is a basic result of
quantum information theory well known as \emph{strong subadditivity
of von Neumann entropy} \cite{Ruskai}. Devetak and Jard established an operational sense of QCMI as a cost of a quantum state
redistribution protocol \cite{D&J}.\smallskip

In infinite dimensions formula (\ref{cmi-d}) may contain the uncertainty $+\infty-\infty$. In \cite{CMI}
it is shown that the r.h.s. of (\ref{cmi-d}) has a unique lower semicontinuous extension to the space of all
states of infinite-dimensional tripartite system possessing all basic properties of QCMI.\smallskip

\begin{theorem}\label{cmi-th} \cite{CMI} \emph{Let $A,B,C$ and $D$ be infinite-dimensional quantum systems.}\smallskip

\noindent A) \emph{There exists  a unique  lower semicontinuous function $I_\mathrm{e}(A\!:\!B|C)_{\rho}$ on the set $\,\S(\H_{ABC})$ taking values in $[0,+\infty]$ such that:
\begin{itemize}
   \item $I_\mathrm{e}(A\!:\!B|C)_{\rho}$ coincides with the r.h.s. of (\ref{cmi-d}) if $H(\rho_{ABC})$ and $H(\rho_{C})$ are finite;
   \item the function $I_\mathrm{e}(A\!:\!B|C)_{\rho}$ possesses the above-stated
properties 1-5 of QCMI.\footnote{The first and the second equalities in property 2 are valid if $I(B\!:\!C)<+\infty$ and $I(A\!:\!C)<+\infty$ correspondingly.}
\end{itemize}}

\noindent B) \emph{The function $I_\mathrm{e}(A\!:\!C|B)_{\rho}$ can be defined by one of the equivalent
expressions\footnote{It is assumed that $I(X\!:\!Y)_{Q\shs\rho Q}=[\Tr
Q\shs\rho]I(X\!:\!Y)_{\frac{Q\shs\rho Q}{\Tr Q\shs\rho}}$.}
\begin{equation}\label{cmi-e+}
I_\mathrm{e}(A\!:\!B|C)_{\rho}=\sup_{P_A}\left[\shs
I(A\!:\!BC)_{Q\rho Q}-I(A\!:\!C)_{Q\rho
Q}\shs\right]\!,\;\,Q=P_A\otimes I_B\otimes I_C,
\end{equation}
\begin{equation}\label{cmi-e++}
I_\mathrm{e}(A\!:\!B|C)_{\rho}=\sup_{P_B}\left[\shs
I(AC\!:\!B)_{Q\rho Q}-I(B\!:\!C)_{Q\rho
Q}\shs\right]\!,\;\,Q=I_A\otimes P_B\otimes I_C,
\end{equation}
where the suprema are over all finite rank projectors
$P_A\in\B(\H_A)$ and $P_B\in\B(\H_B)$.}\smallskip

\noindent C) \emph{For an arbitrary state $\,\rho_{ABCD}$ the
following property is valid:
\begin{equation}\label{t-ext-p++}
I_\mathrm{e}(A\!:\!B|C)_{\rho}=\lim_{k\rightarrow+\infty}\lim_{l\rightarrow+\infty}I_\mathrm{e}(A\!:\!B|C)_{\rho^{klt}}
\end{equation}
for $\,t=k$ and for $\,t=l$, where $\rho_{ABCD}^{klt}$ is the state proportional to the operator
$$
P^k_{A}\otimes
P^k_{B}\otimes P^l_{C}\otimes P^t_{D}\cdot\rho_{ABCD}\cdot P^k_{A}\otimes
P^k_{B}\otimes P^l_{C}\otimes P^t_{D},
$$
$\{P^k_{A}\}_k\subset\B(\H_A)$, $\{P^k_B\}_k\subset\B(\H_{B})$,
$\{P^l_C\}_l\subset\B(\H_C)$ and  $\{P^t_D\}_t\subset\B(\H_D)$  are
sequences of projectors strongly converging to the unit
operators $I_A$,$I_B$,$I_C$ and $I_D$ such that $\,\min\{\rank P^k_A,\rank P^k_B\}<+\infty$ for all $\shs k$.}
\end{theorem}\medskip

We will call the function $I_\mathrm{e}(A\!:\!B|C)_{\rho}$ the (extended) QCMI and will omit the
subscript $\mathrm{e}$. The benefits of using this function is pointed briefly in the Introduction.

The approximation property stated in part C of Theorem \ref{cmi-th} allows to prove
for the extended QCMI the relations valid for the QCMI in the finite-dimensional case. For example,
it allows to show that the following relation
\begin{equation}\label{chain}
I(AD\!:\!BE|C)_{\rho}=I(AD\!:\!E|BC)_{\rho}+I(D\!:\!B|AC)_{\rho}+I(A\!:\!B|C)_{\rho}
\end{equation}
holds for any state $\rho_{ABCDE}$ (with possible values $+\infty$ in both sides).\smallskip

A similar way is used to prove that
\begin{equation}\label{F-c-b}
\left|p
I(A\!:\!B|C)_{\rho}+(1-p)I(A\!:\!B|C)_{\sigma}-I(A\!:\!B|C)_{p\rho+(1-p)\sigma}\right|\leq h_2(p)
\end{equation}
for any $p\in(0,1)$ and any states $\rho_{ABC}$, $\sigma_{ABC}$ with finite $I(A\!:\!B|C)_{\rho}$, $I(A\!:\!B|C)_{\sigma}$, where $h_2(p)$ is the binary entropy \cite{CID}.\smallskip

The lower semicontinuity of $I(A\!:\!B|C)$ on $\S(\H_{ABC})$ means that
$$
\liminf_{n\to +\infty}I(A\!:\!B|C)_{\rho^n}\geq I(A\!:\!B|C)_{\rho^0}
$$
for any sequence $\{\rho^n_{ABC}\}$ converging to a state $\rho^0_{ABC}$. Physically, this
limit relation shows that the conditional total correlation can not be increased by passing
to a limit.  It implies that the set of
all states $\rho_{ABC}$ determined by the inequality
\begin{equation}\label{eps-MC}
I(A\!:\!B|C)_{\rho}\leq \epsilon
\end{equation}
is closed for any $\epsilon\geq0$.  The existence of the Fawzi-Renner recovery channel \cite{F&R}, i.e. a
channel $\Phi:C\rightarrow BC$ such that
\begin{equation}\label{FR-ineq}
F(\rho_{ABC},
\id_A\otimes\Phi(\rho_{AC}))\geq 2^{-\epsilon/2}
\end{equation}
where $F(\varrho,\varsigma)\doteq\|\sqrt{\varrho}\sqrt{\varsigma}\|_1$ is the
quantum fidelity for states $\varrho$ and
$\varsigma$, is proved in \cite{CMI} for \emph{arbitrary} state $\rho_{ABC}$ satisfying (\ref{eps-MC}) by using the approximation technique based on the compactness criterion for families of channels in the strong convergence topology.\footnote{In \cite{F&R,SFR}
the existence of such a channel was proved only for states with finite marginal entropies.}

\smallskip

If either $I(A\!:\!C)_{\rho}$ or $I(B\!:\!C)_{\rho}$ is finite then
expressions (\ref{cmi-e+}) and (\ref{cmi-e++}) are reduced, respectively, to the
following well known formulae
\begin{equation}\label{cmi-d+}
    I(A\!:\!B|C)_{\rho}=I(A\!:\!BC)_{\rho}-I(A\!:\!C)_{\rho},
\end{equation}
\begin{equation}\label{cmi-d++}
    I(A\!:\!B|C)_{\rho}=I(AC\!:\!B)_{\rho}-I(B\!:\!C)_{\rho}.
\end{equation}
These formulae and upper bound (\ref{MI-UB}) imply that
\begin{equation}\label{CMI-UB}
  I(A\!:\!B|C)_{\rho}\leq 2\min\{H(\rho_{A}),H(\rho_{B}),H(\rho_{AC}),H(\rho_{BC})\}.
\end{equation}
By formulae (\ref{cmi-d+}) and (\ref{cmi-d++}) the lower semicontinuity of QCMI,  Lemma \ref{tl} and continuity condition (\ref{mi-cont}) for the quantum  mutual information imply the following continuity conditions for QCMI. \smallskip

\begin{corollary}\label{cmi-th-c-1} \cite{CMI} \emph{If $\{\rho_{ABC}^n\}$ is a sequence  converging to a state $\rho_{ABC}^0$
such that $\lim_{n\rightarrow+\infty}H(\rho^n_{X})=H(\rho^0_{X})<+\infty$, where $X$ is one of the systems
$A,B,AC$ and $BC$ then}
$$
\lim_{n\rightarrow+\infty}I(A\!:\!B|C)_{\rho^n}=I(A\!:\!B|C)_{\rho^0}<+\infty.
$$
\end{corollary}
These conditions will be substantially strengthened in Section 5.1.1 (Corollary \ref{cmi-th-c-1+}).\smallskip

By Theorem \ref{cmi-th} the functions $\rho\mapsto I(AD\!:\!E|BC)_{\rho}$ and $\rho\mapsto I(D\!:\!B|AC)_{\rho}$ are lower semicontinuous
on the set of all states in $\S(\H_{ABCDE})$, where $A,B,C,D$ and $E$ are arbitrary infinite-dimensional systems.
Thus, the relation (\ref{chain}) implies the following \smallskip

\begin{corollary}\label{cmi-th-c-2} \emph{The function $\,\rho\mapsto
\left[\shs I(AD\!:\!BE|C)_{\rho}-I(A\!:\!B|C)_{\rho}\right]$ is lower
semicontinuous on the set
$\,\{\shs\rho\in\S(\H_{ABCDE})\,|\,I(A\!:\!B|C)_{\rho}<+\infty\shs\}$.}
\end{corollary}\smallskip

This corollary will be essentially used in Section 4.

\subsection{Multipartite system}

The quantum conditional mutual information of a state $\,\rho_{A_1 \ldots
A_mC}\,$ of a finite-dimensional multipartite system $A_1 \ldots
A_mC$ is defined as follows (cf.\cite{L-mi,Herbut,NQD,AHS,Y&C})
\begin{equation}\label{cmi-mpd}
\begin{array}{cl}
     I(A_1\!:\ldots:\!A_m|C)_{\rho}&\doteq\displaystyle
    \sum_{k=1}^m H(A_k|C)_{\rho}-H(A_1 \ldots A_m|C)_{\rho}\\&
    =\displaystyle\sum_{k=1}^{m-1} H(A_k|C)_{\rho}-H(A_1 \ldots A_{m-1}|A_mC)_{\rho}.
\end{array}
\end{equation}
Its nonnegativity and other basic properties can be derived from the
corresponding properties of the tripartite QCMI by using the representation (cf.\cite{Y&C})
\begin{equation}\label{cmi-mpd+}
\begin{array}{rl}
     I(A_1\!:\ldots:\!A_m|C)_{\rho}=I(A_2\!:\!A_1|C)_{\rho}\!\!& +\; I(A_3\!:\!A_1A_2|C)_{\rho}+...\\\\&+\;
     I(A_m\!:\!A_1...A_{m-1}|C)_{\rho}.
\end{array}
\end{equation}

By using representation (\ref{cmi-mpd+}) and the extended
QCMI described in Theorem \ref{cmi-th}
one can define QCMI for any state of
an infinite-dimensional system $A_1...A_mC$.

\smallskip

\begin{property}\label{cmi-pr}\cite{CMI} \emph{Let $A_1,...A_m$ and $C$ be infinite-dimensional quantum systems.}\smallskip

\noindent A) \emph{There is a unique lower semicontinuous function  $\,I_{\mathrm{e}}(A_1\!:\!...\!:\!A_m|C)_{\rho}$  on the set
$\,\S(\H_{A_1 \ldots A_mC})$  coinciding with the r.h.s. of (\ref{cmi-mpd}) for any state $\rho_{A_1...A_mC}$ with finite marginal entropies and possessing the analogs of above-stated properties 1-4 of QCMI. This function  can be defined by formula
(\ref{cmi-mpd+}) in which each summand $\,I(X\!:\!Y|C)_{\rho}$
coincides with the function $I_{\mathrm{e}}(X\!:\!Y|C)_{\rho}$
described in Theorem \ref{cmi-th}.}\smallskip

\noindent B) \emph{For an arbitrary state $\,\rho_{A_1...A_mCD}$ the
following property is valid:
\begin{equation}\label{cmi-n-lr}
I_{\mathrm{e}}(A_1\!:\ldots:\!A_m|C)_{\rho}=\lim_{k\rightarrow+\infty}\lim_{l\rightarrow+\infty}I_{\mathrm{e}}(A_1\!:\ldots:\!A_m|C)_{\rho^{klt}}
\end{equation}
for $\,t=k$ and for $\,t=l$, where $\rho_{A_1...A_mCD}^{klt}$ is the state proportional to the operator
$$
P^k_{A_1}\otimes\ldots\otimes P^k_{A_m}\otimes P^l_{C}\otimes P^t_{D}\cdot\rho_{A_1...A_mCD}\cdot P^k_{A_1}\otimes\ldots\otimes P^k_{A_m}\otimes P^l_{C}\otimes P^t_{D},
$$
$\{P^k_{A_i}\}_k\subset\B(\H_{A_i})$, $i=\overline{1,m}$, $\{P^l_C\}_l\subset\B(\H_{C})$ and
$\{P^t_D\}_t\subset\B(\H_{D})$ are sequences of projectors strongly
converging to the unit operators $I_{A_i}$, $i=\overline{1,m}$, $I_{C}$ and $I_D$ such
that\break $\min_{1\leq j \leq m}\sum_{i\neq
j}\mathrm{rank}P^k_{A_i}<+\infty$ for all $\,k$.}\smallskip

\end{property}\medskip

The approximation property stated in part B of Proposition  \ref{cmi-pr} allows to prove
for the function $I_{\mathrm{e}}(A_1\!:\ldots:\!A_m|C)$ the relations valid for the multipartite QCMI in the finite-dimensional case.
It allows to prove, in particular, that representation (\ref{cmi-mpd+}) is valid for
$\,I_{\mathrm{e}}(A_1\!:\ldots:\!A_m|C)_{\rho}$ with arbitrarily permuted
indexes $\,1,...,m\,$ in the right hand side (provided each summand
$I(X\!:\!Y|Z)_{\rho}$ coincides with
$I_{\mathrm{e}}(X\!:\!Y|Z)_{\rho}$) \cite{CMI}.

We will call the function $I_{\mathrm{e}}(A_1\!:\ldots:\!A_m|C)$ the (extended) multipartite QCMI and will omit the
subscript $\mathrm{e}$.

We will use the m-partite version of inequality (\ref{F-c-b}): let $\rho_{A_1...A_mC}$ and $\sigma_{A_1...A_mC}$ be states such that  $R=I(A_1\!:\ldots:\!A_m|C)_{\rho}$ and $S=I(A_1\!:\ldots:\!A_m|C)_{\sigma}$ are finite. Then
\begin{equation}\label{F-c-b+}
-h_2(p)\leq I(A_1\!:\ldots:\!A_m|C)_{p\rho+(1-p)\sigma}-[pR+(1-p)S]\leq (m-1) h_2(p)
\end{equation}
for any $p\in(0,1)$, where $h_2(p)$ is the binary entropy.\smallskip

If $\rho_{A_1...A_mC}$ and $\sigma_{A_1...A_mC}$ are states with finite marginal entropies then inequality (\ref{F-c-b+})
can be proved by using the second expression in (\ref{cmi-mpd}), concavity of the conditional entropy and  inequality (\ref{w-k-ineq}).
The validity of (\ref{F-c-b+}) for arbitrary states $\rho_{A_1...A_mC}$ and $\sigma_{A_1...A_mC}$ with finite QCMI can be shown by
using the approximation property stated in part B of Proposition  \ref{cmi-pr}.

\section{Lower semicontinuity of the loss of QCMI under local channels}

\subsection{Tripartite system}

One of the basic properties of QCMI is the monotonicity under local channels. It means  that
$$
I(A\!:\!B|C)_{\rho}\geq
I(A'\!:\!B'|C)_{\Phi_A\otimes\Phi_B\otimes\id_{\!C}(\rho)}
$$
for any state $\rho_{ABC}$ and arbitrary channels $\Phi_A:A\rightarrow A'$ and $\Phi_B:B\rightarrow
    B'$. So, the quantity
$$
\Delta(\rho_{ABC},\Phi_A,\Phi_B)\doteq I(A\!:\!B|C)_{\rho}-I(A'\!:\!B'|C)_{\Phi_A\otimes\Phi_B\otimes\id_{\!C}(\rho)}
$$
is well defined as a nonnegative number or $+\infty$ for any $\rho_{ABC}$, $\Phi_A$ and $\Phi_B$ such that
$I(A'\!:\!B'|C)_{\Phi_A\otimes\Phi_B\otimes\id_{\!C}(\rho)}<+\infty$. We will call it the loss of QCMI under  local channels.\smallskip

To analyse the action of a local channel on the conditioning system we will consider the quantity
$$
\Delta^{\!\rm c}(\rho_{ABC},\Phi_C)\doteq I(A\!:\!B|C)_{\rho}-I(A\!:\!B|C')_{\id_{AB}\otimes\Phi_{\!C}(\rho)}
$$
which can take values of different sign (in contrast to $\Delta(\rho_{ABC},\Phi_A,\Phi_B)$).\smallskip

We will consider  $\Delta(\rho_{ABC},\Phi_A,\Phi_B)$ as
a function of a triplet $(\rho_{ABC},\Phi_A,\Phi_B)$ and $\Delta^{\!\rm c}(\rho_{ABC},\Phi_C)$ as
a function of a pair $(\rho_{ABC},\Phi_C)$. \smallskip

\begin{theorem}\label{main}
A) \emph{The function $\Delta(\rho_{ABC},\Phi_A,\Phi_B)$ is lower semicontinuous on the set}\footnote{$\F(X,X')$ is the set of all quantum channels from a system $X$ to a system $X'$ equipped with the strong convergence topology (see Section 2).}
$$
\left\{(\rho_{ABC},\Phi_A,\Phi_B)\in\S(\H_{ABC})\times\F(A,A')\times\F(B,B')\,\left|\,I(A'\!:\!B'|C)_{\Phi_A\otimes\Phi_B\otimes\id_{\!C}(\rho)}<+\infty\right\}\right..
$$

\noindent B) \emph{The function $\Delta^{\!\rm c}(\rho_{ABC},\Phi_C)$ is lower semicontinuous and finite on the set
$$
\left\{(\rho_{ABC},\Phi_C)\in\S_{\rm c}\times\F(C,C')\,\left|\,I(A\!:\!B|C)_{\rho}<+\infty\right\}\right.,
$$
where $\S_{\rm c}$ is any subset of $\,\S(\H_{ABC})$ on which the function $\rho_{ABC}\mapsto H(\rho_C)$ is continuous. This function is bounded from below by the value $-2H(\rho_C)$.}\smallskip

\noindent C) \emph{The function $\Delta^{\!\rm c}(\rho_{ABC},\Phi_C)$ is continuous and bounded on the set
$$
\left\{(\rho_{ABC},\Phi_C)\in\S(\H_{ABC})\times\F_k(C,C')\,\left|\,I(A\!:\!B|C)_{\rho}<+\infty\right\}\right.
$$
for each $k\in \mathbb{N}$, where $\F_k(C,C')$ is the subsets of $\F(C,C')$ consisting of
channels with Choi rank not exceeding $k$. The modulus of $\,\Delta^{\!\rm c}(\rho_{ABC},\Phi_C)$ on this set does not exceed $2\log k$.}
\end{theorem}
\medskip

Theorem \ref{main}A states that
\begin{equation}\label{main-rel}
\liminf_{n\to+\infty}\Delta(\rho_{ABC}^n,\Phi^n_A,\Phi^n_B)\geq \Delta(\rho_{ABC}^0,\Phi^0_A,\Phi^0_B)
\end{equation}
for any sequence $\{\rho_{ABC}^n\}$ converging to a state $\rho_{ABC}^0$ and
any sequences $\{\Phi^n_A\}\subset\F(A,A')$ and $\{\Phi^n_B\}\subset\F(B,B')$ strongly converging, respectively, to channels
$\Phi^0_A$ and $\Phi^0_B$ provided that $I(A'\!:\!B'|C)_{\Phi^n_A\otimes\Phi^n_B\otimes\id_{\!C}(\rho^n)}<+\infty$ for
all $n\geq0$ (otherwise the quantity $\Delta(\rho^n_{ABC},\Phi^n_A,\Phi^n_B)$ is not defined).\smallskip

Theorem \ref{main}B states that
\begin{equation}\label{C-ls-lr}
\liminf_{n\to+\infty}\Delta^{\!\rm c}(\rho_{ABC}^n,\Phi^n_C)\geq \Delta^{\!\rm c}(\rho_{ABC}^0,\Phi^0_C)\geq-2H(\rho^0_C)
\end{equation}
for any sequence $\{\rho_{ABC}^n\}$ converging to a state $\rho_{ABC}^0$ and
any sequence $\{\Phi^n_C\}\subset\F(C,C')$ strongly converging to a channel
$\Phi^0_C$ provided that the sequence $\{H(\rho^n_C)\}$
tends to $H(\rho^0_C)<+\infty$ and $I(A\!:\!B|C)_{\rho^n}<+\infty$ for all $n\geq0$. \smallskip

Theorem \ref{main}C states that
\begin{equation}\label{C-ls-lr-2}
\lim_{n\to+\infty}\Delta^{\!\rm c}(\rho_{ABC}^n,\Phi^n_C)=\Delta^{\!\rm c}(\rho_{ABC}^0,\Phi^0_C)\quad \textrm{and}\quad |\Delta^{\!\rm c}(\rho_{ABC}^0,\Phi^0_C)|\leq2\log k
\end{equation}
for any sequence $\{\rho^n_{ABC}\}$ converging to a state $\rho^0_{ABC}$ and
any sequence $\{\Phi^n_C\}\subset\F(C,C')$ strongly converging to a channel
$\Phi^0_C$ provided that the Choi rank of all the channels $\Phi^n_C$ does not exceed $k$ and $I(A\!:\!B|C)_{\rho^n}<+\infty$ for all $n\geq0$.  \smallskip

\emph{Proof.} A) Let $\{\rho_{ABC}^n\}$ be a sequence of states converging to a state $\rho_{ABC}^0$. Let
$\{\Phi^n_A\}$ and $\{\Phi^n_B\}$ be sequences of channels in $\F(A,A')$ and in $\F(B,B')$ strongly converging to channels
$\Phi^0_A$ and $\Phi^0_B$ correspondingly such that $I(A'\!:\!B'|C)_{\Phi^n_A\otimes\Phi^n_{B}\otimes\id_{C}(\rho^n)}<+\infty$ for
all $n\geq0$. By Theorem 2B in \cite{CSR} there exist separable Hilbert spaces $\H_D$ and $\H_E$ and sequences
of isometries $V_n:\H_A\rightarrow\H_{A'D}$ and $U_n:\H_B\rightarrow\H_{B'E}$ strongly converging to  isometries $V_0$ and $U_0$ such that
$$
\Phi^n_A(\varrho)=\Tr_D V_n\varrho V^*_n \quad \textrm{and} \quad \Phi^n_B(\varsigma)=\Tr_E U_n\varsigma U^*_n
$$
for any $\varrho\in\S(\H_A)$, $\varsigma\in\S(\H_B)$ and all $n\geq 0$.

The sequence of states $\sigma^n_{A'B'CDE}=V_n\otimes U_n\otimes I_{C}\cdot\rho_{ABC}^n\cdot V^*_n\otimes U^*_n\otimes I_{C}$  converges to the
state $\sigma_{A'B'CDE}^0=V_0\otimes U_0\otimes I_{C}\cdot\rho_{ABC}^0 \cdot V^*_0\otimes U^*_0\otimes I_{C}$. Since
$$
I(A\!:\!B|C)_{\rho^n}=I(A'D\!:\!B'E|C)_{\sigma^n}\quad \textrm{and} \quad I(A'\!:\!B'|C)_{\Phi^n_A\otimes\Phi^n_B\otimes\id_{C}(\rho^n)}=I(A'\!:\!B'|C)_{\sigma^n}
$$
for all $n\geq0$, relation (\ref{main-rel}) follows from Corollary \ref{cmi-th-c-2} in Section 3. \smallskip

B) Let $\{\rho_{ABC}^n\}$ be a sequence of states converging to a state $\rho_{ABC}^0$ such that\break $I(A\!:\!B|C)_{\rho^n}<+\infty$ for all $n\geq0$ and $\{\Phi^n_C\}$ a sequences of channels in $\F(C,C')$ strongly converging to a channel
$\Phi^0_C$. We have to prove (\ref{C-ls-lr}) assuming that the sequence $\{H(\rho^n_C)\}$ tends to $H(\rho^0_C)<+\infty$.

By Theorem 2B in \cite{CSR} there is a separable Hilbert space $\H_E$ and a sequence
of isometries $V_n:\H_C\rightarrow\H_{C'E}$ strongly converging to an isometry $V_0$ such that
$$
\Phi^n_C(\varrho)=\Tr_E V_n\varrho V^*_n \quad \textrm{for all}\quad \varrho\in\S(\H_C)\textrm{ and }n\geq 0.
$$
The sequence of states $\sigma^n_{ABC'E}=I_{AB}\otimes V_n\cdot\rho^n_{ABC}\cdot I_{AB}\otimes V^*_n$  converges to a
state $\sigma^0_{ABC'E}=I_{AB}\otimes V_0\cdot\rho^0_{ABC}\cdot I_{AB}\otimes V^*_0$. Since
$$
I(A\!:\!B|C)_{\rho^n}=I(A\!:\!B|C'E)_{\sigma^n}\quad \textrm{and} \quad I(A\!:\!B|C')_{\id_{AB}\otimes\Phi^n_{\!C}(\rho^n)}=I(A\!:\!B|C')_{\sigma^n},
$$
Lemma \ref{spl} below implies that
\begin{equation}\label{DC-exp}
\Delta^{\!\rm c}(\rho_{ABC}^n,\Phi^n_C)=I(B\!:\!E|AC')_{\sigma^n}-I(B\!:\!E|C')_{\sigma^n}
\end{equation}
for all $n\geq0$ provided that the last term is finite.

By the assumption we have $\lim_{n\to+\infty}H(\sigma^n_{C'E})=H(\sigma^0_{C'E})<+\infty$,
since $H(\sigma^n_{C'E})=H(V_n\rho^n_CV_n^*)=H(\rho^n_C)$ for all $n\geq0$.
Hence Corollary \ref{cmi-th-c-1} and upper bound (\ref{CMI-UB}) imply that
\begin{equation}\label{ulr+}
\lim_{n\to+\infty}I(B\!:\!E|C')_{\sigma^n}=I(B\!:\!E|C')_{\sigma^0}\leq 2H(\sigma^0_{EC'})=2H(\rho^0_{C}).
\end{equation}
Thus, relation (\ref{C-ls-lr}) follows from (\ref{DC-exp}) and the lower semicontinuity of QCMI.\smallskip

C) By Theorem 2B in \cite{CSR} in this case we can repeat the arguments from the proof of part B with $k$-dimensional
system $E$. Hence the relations
\begin{equation*}\label{ulr}
\lim_{n\to+\infty}I(B\!:\!E|AC')_{\sigma^n}=I(B\!:\!E|AC')_{\sigma^0}<+\infty,
\end{equation*}
and (\ref{ulr+}) hold by Corollary \ref{cmi-th-c-1}. These relations and (\ref{DC-exp}) imply (\ref{C-ls-lr-2}).

The upper bound on $|\Delta^{\!\rm c}(\rho_{ABC},\Phi_C)|$ follows from the expression similar to (\ref{DC-exp}),
since upper bound (\ref{CMI-UB}) implies that
$$
I(B\!:\!E|AC')_{\sigma}, I(B\!:\!E|C')_{\sigma}\leq 2\log\dim\H_E
$$
for any state $\sigma$ in $\S(\H_{ABC'E})$. $\square$
\smallskip

\begin{lemma}\label{spl} \emph{Let $A,B,C$ and $D$ be arbitrary quantum systems. Then the equality
\begin{equation}\label{spl-eq}
I(A\!:\!B|CD)_{\rho}-I(A\!:\!B|C)_{\rho}=I(D\!:\!B|AC)_{\rho}-I(D\!:\!B|C)_{\rho}
\end{equation}
holds for any state $\rho$ in $\S(\H_{ABCD})$ provided  that both sides of
(\ref{spl-eq}) are well defined (do not contain the uncertainty $+\infty-\infty$).}
\end{lemma}\smallskip

\emph{Proof.} Equality (\ref{spl-eq}) is easily verified for any state $\rho$ with finite
marginal entropies (by using formula (\ref{cmi-d}) for QCMI). Its validity for arbitrary state $\rho$
can be shown by using two step approximation.

Let $\rho$ be a state such that $H(\rho_B)<+\infty$. Let $\{\Phi^n_X\}$ be sequences of channels strongly converging to the identity channels $\id_X$, $X=A,C,D$, such that  any channel $\Phi^n_X$ has a finite-dimensional output. The sequence of states $\rho^n=\Phi^n_A\otimes\id_B\otimes\Phi^n_C\otimes\Phi^n_D(\rho)$ having finite
marginal entropies tends to the state $\rho$. Corollary \ref{cmi-th-c-1} implies that
$$
\begin{array}{c}
I(A\!:\!B|CD)_{\rho^n}\rightarrow I(A\!:\!B|CD)_{\rho}, \quad  I(A\!:\!B|C)_{\rho^n}\rightarrow I(A\!:\!B|C)_{\rho}\\\\
I(D\!:\!B|AC)_{\rho^n}\rightarrow I(D\!:\!B|AC)_{\rho}, \quad  I(D\!:\!B|C)_{\rho^n}\rightarrow I(D\!:\!B|C)_{\rho}
\end{array}
$$
as $n\rightarrow+\infty$, where all the limits are finite. So, since equality (\ref{spl-eq}) holds for the state $\rho^n$ for each $n$, it holds for the state $\rho$.

Now the validity of equality (\ref{spl-eq}) for arbitrary state $\rho$ at which both parts of (\ref{spl-eq}) are well defined
can be easily shown by using the approximating property from part C of Theorem \ref{cmi-th}. $\square$.

\subsection{Multipartite system}

The monotonicity of the extended multipartite  QCMI  under local channels means  that
$$
I(A_1\!:\ldots:\!A_m|C)_{\rho}\geq
I(A'_1\!:\ldots:\!A'_m|C)_{\Phi_{A_1}\otimes...\otimes\Phi_{A_m}\otimes\id_{\!C}(\rho)}
$$
for any state $\rho_{A_1...A_mC}$ and arbitrary channels $\Phi_{A_k}:A_k\rightarrow A_k'$, $k=\overline{1,m}$. So, the quantity
$$
\Delta(\rho_{A_1...A_mC},\Phi_{A_1},\ldots,\Phi_{A_m})\doteq I(A_1\!:\ldots:\!A_m|C)_{\rho}-I(A'_1\!:\ldots:\!A'_m|C)_{\Phi_{A_1}\otimes...\otimes\Phi_{A_m}\otimes\id_{\!C}(\rho)}
$$
is well defined as a nonnegative number or $+\infty$ for any $\Phi_{A_1}$,...,$\Phi_{A_m}$ and $\rho_{A_1...A_mC}$ such that
$I(A'_1\!:\ldots:\!A'_m|C)_{\Phi_{A_1}\otimes...\otimes\Phi_{A_m}\otimes\id_{\!C}(\rho)}<+\infty$. We will call it the loss of the multipartite QCMI under local channels.

To analyse the action of a local channel on the conditioning system we will consider the quantity
$$
\Delta^{\!\rm c}(\rho_{A_1...A_mC},\Phi_C)\doteq I(A_1\!:\ldots:\!A_m|C)_{\rho}-I(A_1\!:\ldots:\!A_m|C')_{\id_{A_1...A_m}\otimes\Phi_{C}(\rho)}
$$
which can take values of different sign (in contrast to $\Delta(\rho_{A_1...A_mC},\Phi_{A_1},\ldots,\Phi_{A_m})$).\smallskip

Denote by $\F_l(\{A_k\},\{A'_k\})$ the Cartesian product of the spaces $\F(A_1,A'_1)$,..., $\F(A_m,A'_m)$ of all local channels equipped with the strong convergence topology (see Section 2).
\smallskip

\begin{theorem}\label{main+}
A) \emph{The function $\Delta(\rho_{A_1...A_mC},\Phi_{A_1},\ldots,\Phi_{A_m})$ is lower semicontinuous on the set
$$
\left\{(\rho_{A_1...A_mC},\Phi_{A_1},...,\Phi_{A_m})\in\S(\H_{A_1...A_mC})\times \F_l(\{A_k\},\{A'_k\})\,\left|\,I(A'_1\!:\ldots:\!A'_m|C)_{\rho'}<+\infty\right\}\right.\!,
$$
where $\rho'_{A'_1...A'_mC}=\Phi_{A_1}\otimes...\otimes\Phi_{A_m}\otimes\id_{\!C}(\rho_{A_1...A_mC})$.}\smallskip

\noindent B) \emph{The function $\Delta^{\!\rm c}(\rho_{A_1...A_mC},\Phi_C)$ is lower semicontinuous and finite on the set
$$
\left\{(\rho_{A_1...A_mC},\Phi_C)\in\S_{\rm c}\times\F(C,C')\,\left|\,I(A_1\!:\ldots:\!A_m|C)_{\rho}<+\infty\right\}\right.,
$$
where $\S_{\rm c}$ is any subset of $\,\S(\H_{A_1...A_mC})$ on which the function $\rho_{A_1...A_mC}\mapsto H(\rho_C)$ is continuous. This function is bounded from below by the value $-2(m-1)H(\rho_C)$.}\smallskip

\noindent C) \emph{The function $\Delta^{\!\rm c}(\rho_{A_1...A_mC},\Phi_C)$ is continuous and bounded on the set
$$
\left\{(\rho_{A_1...A_mC},\Phi_C)\in\S(\H_{A_1...A_mC})\times\F_k(C,C')\,\left|\,I(A_1\!:\ldots:\!A_m|C)_{\rho}<+\infty\right\}\right.
$$
for each $k\in \mathbb{N}$, where $\,\F_k(C,C')$ is the subset of $\,\F(C,C')$ consisting of
channels with the Choi rank not exceeding $k$. The modulus of $\,\Delta^{\!\rm c}(\rho_{A_1...A_mC},\Phi_C)$ on this set does not exceed $2(m-1)\log k$.}
\end{theorem}
\smallskip
\emph{Proof.} All the parts of Theorem \ref{main+} can be easily derived from the corresponding parts of Theorem \ref{main} by using
representation (\ref{cmi-mpd+}) $\square$. \smallskip

Theorem \ref{main+}A states that
$$
\liminf_{n\to+\infty}\Delta(\rho_{A_1...A_mC}^n,\Phi^n_{A_1},\ldots,\Phi^n_{A_m})\geq \Delta(\rho_{A_1...A_mC}^0,\Phi^0_{A_1},\ldots,\Phi^0_{A_m})
$$
for arbitrary sequence $\{\rho_{A_1...A_mC}^n\}$ converging to a state $\rho^0_{A_1...A_mC}$ and
any sequences $\{\Phi^n_{A_1}\}\subset\F(A_1,A_1')$,..., $\{\Phi^n_{A_m}\}\subset\F(A_m,A_m')$  strongly converging, respectively, to channels
$\Phi^0_{A_1}$,..., $\Phi^0_{A_m}$ provided that $I(A'_1\!:\ldots:\!A'_m|C)_{\Phi_{A_1}\otimes...\otimes\Phi_{A_m}\otimes\id_{\!C}(\rho^n)}<+\infty$ for all $n\geq 0$.

Theorem \ref{main+}B states that
\begin{equation*}
\liminf_{n\to+\infty}\Delta^{\!\rm c}(\rho_{A_1...A_mC}^n,\Phi^n_C)\geq \Delta^{\!\rm c}(\rho_{A_1...A_mC}^0,\Phi^0_C)\geq-2(m-1)H(\rho^0_C)
\end{equation*}
for arbitrary sequence $\{\rho^n_{A_1...A_mC}\}$ converging to a state $\rho^0_{A_1...A_mC}$ and
any sequence $\{\Phi^n_c\}\subset\F(C,C')$ strongly converging to a channel
$\Phi^0_C$ provided that  the sequence $\{H(\rho^n_C)\}$ tends to $H(\rho^0_C)<+\infty$ and $I(A_1\!:\ldots:\!A_m|C)_{\rho^n}<+\infty$ for all $n\geq0$.  \smallskip

Theorem \ref{main+}C states that
\begin{equation*}
\lim_{n\to+\infty}\Delta^{\!\rm c}(\rho_{A_1...A_mC}^n,\Phi^n_C)=\Delta^{\!\rm c}(\rho_{A_1...A_mC}^0,\Phi^0_C)\quad \textrm{and} \quad |\Delta^{\!\rm c}(\rho_{A_1...A_mC}^0,\Phi^0_C)|\leq 2(m-1)\log  k 
\end{equation*}
for arbitrary  sequence $\{\rho^n_{A_1...A_mC}\}$ converging to a state $\rho^0_{A_1...A_mC}$ and
any sequence $\{\Phi^n_C\}\subset\F(C,C')$ strongly converging to a channel
$\Phi^0_C$ provided that the Choi rank of all the channels $\Phi^n_C$ does not exceed $k$ and $I(A_1\!:\ldots:\!A_m|C)_{\rho^n}<+\infty$ for all $n\geq0$.

\section{Applications}

\subsection{Continuity conditions for QCMI}

\subsubsection{Tripartite system}

Theorem \ref{main} implies, by Lemma \ref{tl}, the following observations that can be interpreted as preserving of local continuity
of QCMI under action of strongly converging sequences of local channels. \smallskip

\begin{property}\label{main-r} \emph{If $\,\{\rho^n_{ABC}\}$ is a sequence of states converging to a state
$\rho^0_{ABC}$ such that $\,\lim_{n\to+\infty}I(A\!:\!B|C)_{\rho^n}=I(A\!:\!B|C)_{\rho^0}<+\infty$ then
$$
\lim_{n\to+\infty}I(A'\!:\!B'|C)_{\Phi^n_A\otimes\Phi^n_B\otimes\id_{\!C}(\rho^n)}=I(A'\!:\!B'|C)_{\Phi^0_A\otimes\Phi^0_B\otimes\id_{\!C}(\rho^0)}
$$
for arbitrary sequences $\{\Phi^n_A\}\subset\F(A,A')$ and $\{\Phi^n_B\}\subset\F(B,B')$ strongly converging to channels
$\Phi^0_A$ and $\Phi^0_B$ correspondingly.}\footnote{$\F(X,X')$ is the set of all quantum channels from a system $X$ to a system $X'$ equipped with the strong convergence topology (see Section 2).}\smallskip

\emph{If, in addition, $\lim_{n\to+\infty}H(\rho^n_C)=H(\rho^0_C)<+\infty$ then
\begin{equation*}
\lim_{n\to+\infty}I(A'\!:\!B'|C')_{\Phi^n_A\otimes\Phi^n_B\otimes\Phi^n_{\!C}(\rho^n)}=I(A'\!:\!B'|C')_{\Phi^0_A\otimes\Phi^0_B\otimes\Phi^0_{\!C}(\rho^0)}
\end{equation*}
for arbitrary  sequences $\{\Phi^n_A\}\subset\F(A,A')$, $\{\Phi^n_B\}\subset\F(B,B')$ and $\{\Phi^n_C\}\subset\F(C,C')$ strongly converging to  channels
$\Phi^0_A$, $\Phi^0_B$ and $\Phi^0_C$.  The condition $\lim_{n\to+\infty}H(\rho^n_C)=H(\rho^0_C)<+\infty$
can be omitted if all the channels $\Phi_C^n$ have bounded Choi rank.}  \smallskip
\end{property}

To illustrate Proposition \ref{main-r} consider the following \smallskip

\begin{example}\label{e-0}
Let $\rho_{ABC}$ be a state such that $I(A\!:\!B|C)_{\rho}$ and
$H(\rho_C)$ are finite. Let $\{\Phi_A^t:A\rightarrow A\}_{t\in \mathbb{R}_+}$, $\{\Phi_B^t:B\rightarrow B\}_{t\in \mathbb{R}_+}$ and $\{\Phi_C^t:C\rightarrow C\}_{t\in \mathbb{R}_+}$
be arbitrary strongly continuous families of quantum channels (for instance, quantum dynamical semigroups). Proposition \ref{main-r} and Theorem \ref{main} imply that
the function
$$
t\mapsto I(A\!:\!B|C)_{\Phi^t_A\otimes\Phi^t_B\otimes\Phi^t_{\!C}(\rho)}
$$
is continuous on $\mathbb{R}_+$ and bounded above by $I(A\!:\!B|C)_{\rho}+2H(\rho_C)$.
\end{example}

By using Corollary \ref{cmi-th-c-1} and taking partial traces in the role of the channels $\Phi^n_A$ and $\Phi^n_B$ we obtain from Proposition \ref{main-r}
the following condition for local continuity of QCMI.

\smallskip

\begin{corollary}\label{cmi-th-c-1+} \emph{Let $\{\rho_{ABC}^n\}$ be a sequence of states converging to a state $\rho_{ABC}^0$.
If there exist a  system $E$ and a sequence $\{\hat{\rho}_{ABCE}^n\}$ of states converging to a state $\hat{\rho}_{ABCE}^0$ such that
$\hat{\rho}_{ABC}^n=\rho_{ABC}^n$ for all $\,n\geq 0$ and $\,\lim_{n\rightarrow\infty}H(\rho^n_{X})=H(\rho^0_{X})<+\infty$, where $X$ is one of the systems $AE$, $BE$, $ACE$ and  $BCE$, then}
$$
\lim_{n\rightarrow\infty}I(A\!:\!B|C)_{\rho^n}=I(A\!:\!B|C)_{\rho^0}<+\infty.
$$
\end{corollary}

The conditions for local continuity of QCMI given by Corollary \ref{cmi-th-c-1+} is more powerful than the conditions
given by Corollary \ref{cmi-th-c-1} (which corresponds to Corollary \ref{cmi-th-c-1+} with trivial system $E$). To show this it suffices to consider the case $I(A\!:\!B|C)=I(A\!:\!B)$. In this case
Corollary \ref{cmi-th-c-1+} states that $I(A\!:\!B)_{\rho^n}$ tends to $I(A\!:\!B)_{\rho^0}$ for a sequence $\{\rho_{AB}^n\}$ converging to a state $\rho_{AB}^0$ provided that
\begin{equation}\label{mi-cont+}
\!\exists \{\hat{\rho}^n_{ABE}\}\rightarrow \hat{\rho}^0_{ABE}\quad \textup{such that} \quad  \hat{\rho}^n_{AB}=\rho^n_{AB}\;\, \forall n\geq0\quad \textrm{and}\quad  H(\hat{\rho}^n_{X})\rightarrow H(\hat{\rho}^0_{X}),
\end{equation}
where $X$ is either $AE$ or $BE$. So, to prove continuity of $I(A\!:\!B)$ for a given converging sequence $\{\rho^n_{AB}\}$ it suffices, briefly speaking,  to find a converging sequence $\{\hat{\rho}^n_{ABE}\}$ of extensions such that either $H(\hat{\rho}^n_{AE})$ or $H(\hat{\rho}^n_{BE})$ is continuous.
The following example demonstrates how to use this condition.\smallskip

\begin{example}\label{e-1}
Let
$$
\rho^n_{AB}=\sum_{i=1}^{d}p_i^n\shs\alpha_i^n\otimes \beta_i^n, \quad n=0,1,2,...,
$$
where $d\leq+\infty$, $\alpha_i^n\in\S(\H_A)$ and $\beta_i^n\in\S(\H_B)$ for all $i$ and $n$ and $\{p_i^n\}_{i=1}^{d}$ is a probability
distribution for each $n\geq0$. Assume that
$$
\lim_{n\to+\infty}\alpha_i^n=\alpha_i^0,\quad \lim_{n\to+\infty}\beta_i^n=\beta_i^0\quad \textrm{and} \quad\lim_{n\to+\infty}p_i^n=p_i^0\quad \textrm{for each}\; i.
$$
By using Lemma \ref{D-A} and its classical counterpart it is easy to show that the sequence $\{\rho^n_{AB}\}$ tends to the state $\rho^0_{AB}$.

Let $\H_E=\bigoplus_{i=1}^{d}\H_{E_i}$, where $\H_{E_i}$ is a separable Hilbert space for any $i$. For each $i$ let $\{\hat{\alpha}_i^n\}_n$ be a sequence of pure states in $\H_{AE_i}$ converging to a pure state $\hat{\alpha}_i^0$ such that $\Tr_{E_i}\hat{\alpha}_i^n=\alpha_i^n$ for all $n\geq0$.
By using Lemma \ref{D-A} it is easy to show that
$$
\hat{\rho}^n_{ABE}=\sum_{i=1}^{d}p_i^n\shs \hat{\alpha}_i^n\otimes \beta_i^n\quad\rightarrow\quad\hat{\rho}^0_{ABE}=\sum_{i=1}^{d}p_i^0\shs \hat{\alpha}_i^0\otimes \beta_i^0 \quad \textrm{as} \quad n\rightarrow+\infty.
$$
It is clear that $\hat{\rho}^n_{AB}=\rho^n_{AB}$  for all $n\geq0$. Since for each $n$ all the pure states $\hat{\alpha}_1^n $, $\hat{\alpha}_2^n $, ... are mutually
orthogonal, $H(\hat{\rho}^n_{AE})=S(\{p^n_i\})$ -- the Shannon entropy of the probability distribution $\{p^n_i\}$. Thus, condition
(\ref{mi-cont+}) implies that
$$
\lim_{n\rightarrow+\infty}I(A\!:\!B)_{\rho^n}=I(A\!:\!B)_{\rho^0}<+\infty\quad \textrm{provided that}\quad \lim_{n\rightarrow+\infty}S(\{p^n_i\})=S(\{p^0_i\})<+\infty.
$$
In particular, if $d<+\infty$ then $\,I(A\!:\!B)_{\rho^n}\,$ always tends to $\,I(A\!:\!B)_{\rho^0}$.

It looks natural (keeping in mind the physical sense of $I(A\!:\!B)$) that the above continuity condition for $I(A\!:\!B)_{\rho^n}$ does not depend on properties of the states
$\alpha_i^n$ and $\beta_i^n$, in particular, on their entropies. Note that this condition can not be obtained
from Corollary \ref{cmi-th-c-1}, i.e. without construction of appropriate extension of the sequence $\{\rho^n_{AB}\}$. Moreover, if $H(\alpha_{i_1}^0)=H(\beta_{i_2}^0)=+\infty$ for some $i_1$ and $i_2$ then
Corollary \ref{cmi-th-c-1} can say nothing about the convergence of $I(A\!:\!B)_{\rho^n}$ to $I(A\!:\!B)_{\rho^0}$ for the above sequence of states. $\square$
\end{example}\smallskip

Since we have not found a sequence $\{\rho^n_{AB}\}$ converging to a state $\rho^0_{AB}$ such that
\begin{equation}\label{mi-conv}
\lim_{n\rightarrow+\infty}I(A\!:\!B)_{\rho^n}=I(A\!:\!B)_{\rho^0}<+\infty
\end{equation}
but condition (\ref{mi-cont+}) is not valid, it is reasonable to propose the following\smallskip

\textbf{Conjecture.} Condition
(\ref{mi-cont+}) is equivalent to (\ref{mi-conv}).\footnote{I would be grateful for any comments concerning this conjecture.}  \smallskip

To prove this conjecture it is sufficient to show that
for any sequence $\{\rho^n_{AB}\}$ converging to a state $\rho^0_{AB}$ such that
(\ref{mi-conv}) holds there is a sequence $\{\hat{\rho}^n_{AA'BB'}\}$ of pure states converging to a pure state $\hat{\rho}^0_{AA'BB'}$
such that
$$
\lim_{n\rightarrow+\infty}I(AA'\!:\!BB')_{\hat{\rho}^n}=I(AA'\!:\!BB')_{\hat{\rho}^0}<+\infty.
$$

\subsubsection{Multipartite system}

The following multipartite version of Proposition \ref{main-r} is obtained from Theorem \ref{main+} by means of Lemma \ref{tl}.
\smallskip

\begin{property}\label{main-r+} \emph{If $\,\{\rho^n_{A_1...A_mC}\}$ is a sequence of states converging to a state
$\rho^0_{A_1...A_mC}$ such that
\begin{equation}\label{mmi-c}
\lim_{n\to+\infty}I(A_1\!:\ldots:\!A_m|C)_{\rho^n}=I(A_1\!:\ldots:\!A_m|C)_{\rho^0}<+\infty
\end{equation}
then
$$
\lim_{n\to+\infty}I(A'_1\!:\ldots:\!A'_m|C)_{\Phi^n_{A_1}\otimes...\otimes\Phi^n_{A_m}\!\otimes\id_{\!C}(\rho^n)}
=I(A'_1\!:\ldots:\!A'_m|C)_{\Phi^0_{A_1}\otimes...\otimes\Phi^0_{A_m}\!\otimes\id_{\!C}(\rho^0)}
$$
for arbitrary sequences $\{\Phi^n_{A_1}\}\subset\F(A_1,A_1')$,...,$\{\Phi^n_{A_m}\}\subset\F(A_m,A_m')$  strongly converging to channels
$\Phi^0_{A_1}$,..., $\Phi^0_{A_m}$ correspondingly.}\smallskip

\emph{If, in addition, $\lim_{n\to+\infty}H(\rho^n_C)=H(\rho^0_C)<+\infty$ then
\begin{equation*}
\lim_{n\to+\infty}I(A'_1\!:\ldots:\!A'_m|C')_{\Phi^n_{A_1}\otimes...\otimes\Phi^n_{A_m}\!\otimes\Phi^n_{\!C}(\rho^n)}
=I(A'_1\!:\ldots:\!A'_m|C')_{\Phi^0_{A_1}\otimes...\otimes\Phi^0_{A_m}\!\otimes\Phi^0_{\!C}(\rho^0)}
\end{equation*}
for arbitrary  sequences $\{\Phi^n_{A_1}\}\subset\F(A_1,A_1')$,...,$\{\Phi^n_{A_m}\}\subset\F(A_m,A_m')$ and $\{\Phi^n_C\}\subset\F(C,C')$ strongly converging to  channels $\,\Phi^0_{A_1}$,..., $\Phi^0_{A_m}$ and $\Phi^0_C$ correspondingly.  The condition $\,\lim_{n\to+\infty}H(\rho^n_C)=H(\rho^0_C)<+\infty$
can be omitted if all the channels $\Phi_C^n$ have bounded Choi rank.}  \smallskip
\end{property}

By Proposition 5D in \cite{CMI} relation (\ref{mmi-c}) holds if
there is a set of indexes $i_1,...,i_{m-1}$  s.t.
$\,\lim_{n\rightarrow\infty}H(\rho^n_{X_{i_k}})=H(\rho^0_{X_{i_k}})<+\infty$, $k=\overline{1,m-1}$,
where $X_{i_k}$ is either $A_{i_k}$ or $A_{i_k}C$.

\smallskip

By taking partial traces in the role of the channels $\Phi^n_{A_i}$ in Proposition \ref{main-r+} one can
strengthen the above condition as follows.
\smallskip

\begin{corollary}\label{main-r+c} \emph{Let $\{\rho^n_{A_1...A_mC}\}$ be a sequence of states converging to a state
$\rho^0_{A_1...A_mC}$. If there exist systems $E_1,...E_{m-1}$ and a sequence $\{\hat{\rho}_{A_1...A_mCE_1...E_{m-1}}^n\}$ of states converging to a state $\hat{\rho}_{A_1...A_mCE_1...E_{m-1}}^0$ such that
$\hat{\rho}_{A_1...A_mC}^n=\rho_{A_1...A_mC}^n$ for all $\,n\geq 0$ and $\,\lim_{n\rightarrow+\infty}H(\rho^n_{X_{i_k}})=H(\rho^0_{X_{i_k}})<+\infty$ for some set of indexes $i_1,...,i_{m-1}$,  where $X_{i_k}$ is either $A_{i_k}E_{k}$ or $A_{i_k}CE_{k}$, for each $k=\overline{1,m-1}$, then relation  (\ref{mmi-c}) holds.}
\end{corollary}\smallskip

Corollary \ref{main-r+c} is a $m$-partite version of Corollary \ref{cmi-th-c-1+}. It allows to prove relation  (\ref{mmi-c}) for a converging sequence
of states by constructing appropriate extension of this sequence.\smallskip

Let $\{\rho^n_{A_1...A_m}\}$ be a natural $m$-partite generalization of the converging sequence of separable states
from the above Example 1. By the similar way one can construct a sequence $\{\hat{\rho}_{A_1...A_mE_1...E_{m-1}}^n\}$ of extensions
converging to an extension $\{\hat{\rho}_{A_1...A_mE_1...E_{m-1}}^0\}$ of the state $\rho^0_{A_1...A_m}$
such that for $k=\overline{1,m-1}$ and all $n\geq0$ the von Neumann entropy $H(\rho^n_{A_{k}E_k})$ coincides with the Shannon entropy $S(\{p^n_i\})$ of the probability distribution from the decomposition of $\rho^n_{A_1...A_m}$ into product states. Thus, Corollary \ref{main-r+c}  allows to show that
$$
\lim_{n\rightarrow+\infty}S(\{p^n_i\})=S(\{p^0_i\})<+\infty\quad\Rightarrow\quad\lim_{n\to+\infty}I(A_1\!:\ldots:\!A_m)_{\rho^n}=I(A_1\!:\ldots:\!A_m)_{\rho^0}
$$
\emph{regardless} of characteristics of the product states involved in the decompositions of the states $\rho^n_{A_1...A_m}$. Note that this assertion
can not be proved by using the conditions for local continuity of $\,I(A_1\!:\ldots:\!A_m)$ existing in the literature
(in particular, by using Proposition 5D in \cite{CMI} mentioned before).

\subsection{Continuity conditions for the squashed entanglement}

\subsubsection{Bipartite system}

The squashed entanglement is one of the basic entanglement measures defined for a state $\rho_{AB}$ of a finite-dimensional bipartite system $AB$ as
\begin{equation}\label{se-def}
  E_{sq}(\rho_{AB})=\textstyle\frac{1}{2}\displaystyle\inf_{\rho_{ABE}}I(A\!:\!B|E),
\end{equation}
where the infimum is over all extensions $\rho_{ABE}$ of the state
$\rho_{AB}$ \cite{C&W,Tucci}. By using the extended QCMI defined by the equivalent formulae (\ref{cmi-d+}) and (\ref{cmi-d++})
this definition can be generalized for any state $\rho_{AB}$ of an infinite-dimensional bipartite system $AB$. It is shown in \cite{SE} that in this case the function $E_{sq}$ defined in
(\ref{se-def}) possesses all basic properties of an entanglement measure (cf.\cite{H,P&V}) excepting the global continuity  and the vanishing  on the set of all separable states (the latter property is conjectured but not proved because of the existence of separable states that are not countably-decomposable, see Remark 10 in \cite{SE}). It is also shown that the function $E_{sq}$ is lower semicontinuous on the set
$$
\S_\mathrm{*}(\H_{AB})\doteq\left\{\rho_{AB}\,|\,\min\{H(\rho_{A}),H(\rho_{B}),H(\rho_{AB})\right\}<+\infty\}
$$
and coincides on this set with the convex closure of $E_{sq}$ -- the maximal convex lower semicontinuous function on $\S(\H_{AB})$  not exceeding $E_{sq}$.\footnote{It is shown in \cite{SE} that the convex closure of $E_{sq}$ can be obtained from the finite-dimensional squashed entanglement by
the construction called universal extension. These is a conjecture that $E_{sq}$ coincides with its convex closure on the whole space $\S(\H_{AB})$.}
This implies, in particular, that $E_{sq}(\rho_{AB})=0$ for any separable state $\rho_{AB}$ belonging to the set $\S_\mathrm{*}(\H_{AB})$ \cite{SE}.

A special attention is paid in  \cite{SE} to analysis of local continuity of $E_{sq}$. It is proved that
$\lim_{n\to+\infty}E_{sq}(\rho^n_{AB})=E_{sq}(\rho^0_{AB})$ for a sequence $\{\rho^n_{AB}\}$ converging
to a state $\rho^0_{AB}$ provided that
\begin{equation}\label{SE-c-1}
\lim_{n\to+\infty}I(A\!:\!B)_{\rho^n}=I(A\!:\!B)_{\rho^0}<+\infty\quad \textrm{and}\quad \min\{H(\rho^0_{A}),H(\rho^0_{B})\}<+\infty.
\end{equation}
By Theorem 1 in \cite{CMI} this condition holds if either
\begin{equation}\label{SE-c-2}
\lim_{n\to+\infty}H(\rho^n_{A})=H(\rho^0_{A})<+\infty\quad \textrm{or}\quad \lim_{n\to+\infty}H(\rho^n_{B})=H(\rho^0_{B})<+\infty.
\end{equation}

By combining Proposition \ref{main-r} in Section 5.1.1 with Proposition 13 in \cite{SE} we obtain the following\smallskip

\begin{property}\label{SE-c} \emph{Let $\{\rho^n_{AB}\}$ be a sequence of states converging to a state $\rho^0_{AB}$.  Then
$$
\lim_{n\to+\infty}I(A\!:\!B)_{\rho^n}=I(A\!:\!B)_{\rho^0}<+\infty\;\;\Rightarrow\;\;
\lim_{n\to+\infty}E_{sq}(\Phi^n_A\otimes\Phi^n_B(\rho^n_{AB}))=E_{sq}(\Phi^0_A\otimes\Phi^0_B(\rho^0_{AB}))
$$
for arbitrary  sequences $\{\Phi^n_A\}\subset\F(A,A')$ and $\{\Phi^n_B\}\subset\F(B,B')$ strongly converging to  channels
$\Phi^0_A$ and $\Phi^0_B$  provided that either $H(\Phi^0_A(\rho^0_{A}))$ or $H(\Phi^0_B(\rho^0_{B}))$ is finite.}\footnote{$\F(X,X')$ is the set of all quantum channels from a system $X$ to a system $X'$ equipped with the strong convergence topology (see Section 2).}
\smallskip
\end{property}

By using Corollary \ref{cmi-th-c-1} and  taking partial traces in the role of the  channels $\Phi^n_A$ and $\Phi^n_B$ we obtain from Proposition \ref{SE-c}
the following condition for local continuity of $E_{sq}$.
\smallskip

\begin{corollary}\label{SE-c+} \emph{Let $\{\rho_{AB}^n\}$ be a sequence  converging to a state $\rho_{AB}^0$ such that either $H(\rho^0_{A})$ or $H(\rho^0_{B})$ is finite. If there exist a system $E$ and a sequence $\{\hat{\rho}_{ABE}^n\}$ converging to a state $\hat{\rho}_{ABE}^0$ such that
$\hat{\rho}_{AB}^n=\rho_{AB}^n$ for all $\,n\geq 0$ and $\,\lim\limits_{n\rightarrow+\infty}H(\rho^n_{X})=H(\rho^0_{X})<+\infty$, where $X$ is either $AE$ or $BE$, then}
$$
\lim_{n\rightarrow+\infty}E_{sq}(\rho_{AB}^n)=E_{sq}(\rho_{AB}^0)<+\infty.
$$
\end{corollary}

The assertion of Corollary \ref{SE-c+} can be also obtained by combining condition (\ref{SE-c-1}) with the continuity
condition (\ref{mi-cont+}) for the quantum mutual information.

The continuity condition given by Corollary \ref{SE-c+} can be treated as a strengthened form of condition (\ref{SE-c-2}). To show superiority (and applicability) of the former condition consider the following\smallskip

\begin{example}\label{e-2}
Let $\{\rho^n_{A}\}$, $\{\rho^n_{B}\}$ and $\{\sigma^n_{AB}\}$ be sequences of states converging, respectively, to states
$\rho^0_{A}$, $\rho^0_{B}$ and $\sigma^0_{AB}$.  Corollary \ref{SE-c+} allows to show that
$$
\lim_{n\rightarrow+\infty} E_{sq}((1-p)\rho^n_{A}\otimes \rho^n_{B}+p\,\sigma^n_{AB})=E_{sq}((1-p) \rho^0_{A}\otimes \rho^0_{B}+p\,\sigma^0_{AB}),\quad p\in[0,1],
$$
provided that
\begin{equation}\label{s-cond}
\lim_{n\rightarrow+\infty}H(\sigma^n_X)=H(\sigma^0_X)<+\infty
\end{equation}
and $H(\rho^0_{X})<+\infty$, where $X$ is either $A$ or $B$, \emph{regardless} of the entropies of the states $\rho^n_{A}$ and $\rho^n_{B}$. Indeed, assume that $X=A$ and $\{\hat{\rho}^n_{AE}\}$ is a sequence of pure states
converging to a pure state $\hat{\rho}^0_{AE}$ such that $\hat{\rho}^n_{A}=\rho^n_{A}$ for all $n\geq0$. Let $\tau_E$ be a pure state orthogonal to all the states $\hat{\rho}^n_{E}$ (we may always extend the system $E$ if necessary). Then the sequence of states
$$
\varrho^n_{ABE}=(1-p)\hat{\rho}^n_{AE}\otimes \rho^n_{B}+p\,\sigma^n_{AB}\otimes \tau_E
$$
tends to the state $\varrho^0_{ABE}=(1-p)\hat{\rho}^0_{AE}\otimes \rho^0_{B}+p\shs\sigma^0_{AB}\otimes \tau_E$, $\varrho^n_{AB}=(1-p)\rho^n_{A}\otimes \rho^n_{B}+p\shs\sigma^n_{AB}$ and
$H(\varrho^n_{AE})=pH(\sigma^n_A)+h_2(p)$ for all $n\geq0$. Thus, the assumption (\ref{s-cond}) with $X=A$ implies that
$$
\lim_{n\rightarrow+\infty} H(\varrho^n_{AE})=H(\varrho^0_{AE})<+\infty.
$$
By considering the case when $p=1$ and all the states $\sigma^n_{AB}$ are pure we see that condition (\ref{s-cond}) is essential. Note that the use of condition (\ref{SE-c-2})
for the above sequence of state leads to the additional "nonphysical" requirement
$$
\lim_{n\rightarrow+\infty}H(\rho^n_X)=H(\rho^0_X)<+\infty.
$$
The above example can be generalized by considering the sequences of the form
$$
(1-p)\rho^n_{AB}+p\,\sigma^n_{AB},
$$
where $\{\rho^n_{AB}\}$ is the converging sequence of separable states from Example \ref{e-1}.
\end{example}

\subsubsection{Multipartite system}

The squashed entanglement of a state $\rho_{A_1...A_m}$ of a  finite-dimensional composite system $A_1...A_m$ is defined as
\begin{equation}\label{mse-def}
  E_{sq}(\rho_{A_1...A_m})=\textstyle\frac{1}{2}\displaystyle\inf_{\rho_{A_1...A_mE}}I(A_1\!:...:\!A_m|E)_{\rho},
\end{equation}
where the infimum is over all extensions $\rho_{A_1...A_mE}$ of the state
$\rho_{A_1...A_m}$ \cite{AHS,Y&C}.\footnote{In \cite{AHS,Y&C} two $m$-partite generalizations of the bipartite squashed entanglement
are proposed: the first one is defined in (\ref{mse-def}), the second one is defined by the expression similar to (\ref{mse-def}) with
the different $m$-partite version of QCMI (called dual conditional total correlation). In \cite{D-S-W} it is proved that these $m$-partite
generalizations of the bipartite squashed entanglement coincide.} By using the extended multipartite QCMI described in Section 3.2
this definition can be generalized for any state $\rho_{A_1...A_m}$ of an infinite-dimensional composite system $A_1...A_m$. By using the arguments from \cite{AHS,Y&C} one can show that in this case the function $E_{sq}$
defined by formula (\ref{mse-def}) possesses almost all properties of an entanglement measure, in particular,
it is convex on the whole set of states of an infinite-dimensional composite system $A_1...A_m$ and nonincreasing under LOCC. Similar to the bipartite case,
it is not clear how to show that $E_{sq}$ is equal to zero on the set of all separable states because of
the existence of countably nondecomposable separable states in infinite-dimensional composite systems (see Remark 10 in \cite{SE}).
Below we show that $E_{sq}$ is equal to zero on the set of all separable states in $\S(\H_{A_1...A_m})$ with finite
marginal entropies  (Corollary \ref{mSE-0}) by using the following proposition describing continuity properties of $E_{sq}$ in infinite dimensions. \smallskip

\begin{property}\label{SE-m} A) \emph{The function $E_{sq}(\rho_{A_1...A_m})$ is lower semicontinuous on the set}
\begin{equation}\label{Star}
\S_*(\H_{A_1...A_m})\doteq\left\{\rho_{A_1...A_m}\,|\,H(\rho_{A_1}),...,H(\rho_{A_m})<+\infty\right\}.
\end{equation}

\noindent B) \emph{If  $\{\rho^n_{A_1...A_m}\}$ is a sequence of states converging to a state $\rho^0_{A_1...A_m}$ such that
\begin{equation}\label{mI-cont}
\lim_{n\to+\infty}I(A_1\!:...:\!A_m)_{\rho^n}=I(A_1\!:...:\!A_m)_{\rho^0}<+\infty
\end{equation}
and $\rho^0_{A_1...A_m}\in\S_*(\H_{A_1...A_m})$ then
\begin{equation}\label{mSE-cont}
\lim_{n\to+\infty}E_{sq}(\rho^n_{A_1...A_m})=
E_{sq}(\rho^0_{A_1...A_m}).
\end{equation}
Moreover, for arbitrary  sequences $\,\{\Phi^n_{A_1}\}\subset\F(A_1,A_1')$,..., $\{\Phi^n_{A_m}\}\subset\F(A_m,A_m')$ strongly converging to channels
$\,\Phi^0_{A_1}$,..., $\Phi^0_{A_m}$  such that all the states $\,\Phi^0_{A_1}(\rho^0_{A_1})$,..., $\Phi^0_{A_m}(\rho^0_{A_m})$ have finite entropy
condition (\ref{mI-cont}) implies that}
$$
\lim_{n\to+\infty}E_{sq}(\Phi^n_{A_1}\otimes...\otimes\Phi^n_{A_m}(\rho^n_{A_1...A_m}))=
E_{sq}(\Phi^0_{A_1}\otimes...\otimes\Phi^0_{A_m}(\rho^0_{A_1...A_m})).
$$

\end{property}

\emph{Proof.} A) Let $\{P_1^n\}\subset \B(\H_{A_1})$,..., $\{P_m^n\}\subset \B(\H_{A_m})$ be sequences of finite rank projectors strongly converging
to the unit operators $I_{A_1}$,..., $I_{A_m}$ such that $\rank P_k^n\leq n$, $k=\overline{1,m}$.  For each $k$ and $n$ consider the channel
\begin{equation}\label{phi-k}
\Phi_k^n(\varrho)=P_k^n\varrho P_k^n+\tau^n_k[\Tr(I_{A_k}-P_k^n)\varrho\shs],
\end{equation}
from $\S(\H_{A_k})$ to itself, where $\tau^n_k$ is a pure state supported by the range of $P_k^n$. It is easy to see that the sequence
$\{\Phi_k^n\}_n$ strongly converges to the identity channel $\id_{A_k}$.

For given $n$ let
$$
f_n(\rho_{A_1...A_m})=\textstyle\frac{1}{2}\displaystyle\inf_{\rho_{A_1...A_mE}}I(A_1\!:...:\!A_m|E)_{\Phi^n_1\otimes...\otimes\Phi^n_m\otimes\id_{\!E}(\rho)}
$$
where the infimum is over all extensions $\rho_{A_1...A_mE}$ of the state
$\rho_{A_1...A_m}$. We will show that the function $f_n$ is continuous on the space $\S(\H_{A_1,.,A_m})$ for each $n$.

Let $\rho^1_{A_1...A_m}$ and  $\rho^2_{A_1...A_m}$ be states such that $\|\rho^1_{A_1...A_m}-\rho^2_{A_1...A_m}\|_1\leq\varepsilon$.
By the arguments from the proof of continuity of $E_{sq}$ in \cite{C&W} we have
$$
f_n(\rho^t_{A_1...A_m})=\textstyle\frac{1}{2}\displaystyle\inf_{\Lambda}I(A_1\!:...:\!A_m|E)_{\Phi^n_1\otimes...\otimes\Phi^n_m\otimes\id_{\!E}(\Lambda(\hat{\rho}^t))},
\quad t=1,2,
$$
where $\hat{\rho}^1_{A_1...A_mR}$ and $\hat{\rho}^2_{A_1...A_mR}$ are purifications
of the states $\rho^1_{A_1...A_m}$ and  $\rho^2_{A_1...A_m}$ such that
$$
\|\hat{\rho}^1_{A_1...A_mR}-\hat{\rho}^2_{A_1...A_mR}\|_1\leq 2\sqrt{\varepsilon}
$$
and the infimum is over all "squashing" channels $\,\Lambda:R\rightarrow E$.

For a given channel $\Lambda$ let $\sigma^t_{A_1...A_mE}=\Phi^n_1\otimes...\otimes\Phi^n_m\otimes\id_E(\Lambda(\hat{\rho}_{A_1...A_mR}^t))$, $t=1,2$.
By monotonicity  of the trace norm under action of a channel we have
$$
\|\sigma^1_{A_1...A_mE}-\sigma^2_{A_1...A_mE}\|_1\leq 2\sqrt{\varepsilon}
$$
Since $\rank\sigma^1_{A_k},\rank\sigma^2_{A_k}\leq n$ for all $k$, Proposition 5E in \cite{CMI} implies that
$$
|I(A_1\!:...:\!A_m|E)_{\sigma^1}-I(A_1\!:...:\!A_m|E)_{\sigma^2}|\leq 2(m-1)\sqrt{\varepsilon}\log n+2mg(\sqrt{\varepsilon}),
$$
where $g(x)=(x+1)\log(x+1)-x\log x$. It follows that $\,|f_n(\rho^1_{A_1...A_m})-f_n(\rho^2_{A_1...A_m})|\leq 2(m-1)\sqrt{\varepsilon}\log n+2mg(\sqrt{\varepsilon})$, and hence the
function $f_n$ is uniformly continuous on $\S(\H_{A_1,.,A_m})$.

Thus, since $f_n(\rho_{A_1...A_m})\leq E_{sq}(\rho_{A_1...A_m})$ for any $\rho_{A_1...A_m}$ and all $n$ by the monotonicity of QCMI, to prove the lower semicontinuity of $E_{sq}$ on the set $\S_*(\H_{A_1...A_m})$ it suffices to show that
$f_n(\rho_{A_1...A_m})$ tends to $E_{sq}(\rho_{A_1...A_m})$ as $n\to+\infty$ for any state $\rho_{A_1...A_m}$ in $\S_*(\H_{A_1...A_m})$.
For a given extension $\rho_{A_1...A_mE}$ of this state we have
\begin{equation}\label{delta-I}
\begin{array}{c}
|I(A_1\!:...:\!A_m|E)_{\rho}-I(A_1\!:...:\!A_m|E)_{\Phi^n_1\otimes...\otimes\Phi^n_m\otimes\id_{\!E}(\rho)}|
\\\\\displaystyle\leq \sum_{k=1}^m|I(A_1\!:...:\!A_m|E)_{\Psi^n_{k-1}\otimes\id_{\!E}(\rho)}-I(A_1\!:...:\!A_m|E)_{\Psi^n_k\otimes\id_{\!E}(\rho)}|,
\end{array}
\end{equation}
where $\Psi^n_k=\Phi^n_1\otimes...\otimes\Phi^n_k\otimes\id_{A_{k+1}..A_{m}}$, $k=\overline{1,m}$, and $\Psi^n_0=\id_{A_{1}..A_{m}}$.
Lemma \ref{imp-l} below implies that
$$
\begin{array}{c}
|I(A_1\!:...:\!A_m|E)_{\Psi^n_{k-1}\otimes\id_{\!E}(\rho)}-I(A_1\!:...:\!A_m|E)_{\Psi^n_k\otimes\id_{\!E}(\rho)}|\\\\\leq
2[H(\rho_{A_k})-H(P_k^n\rho_{A_k}P_k^n)]+h_2(\Tr P_k^n\rho_{A_k}),
\end{array}
$$
since  $[\Psi^n_{k-1}\otimes\id_E(\rho_{A_1...A_m})]_{A_k}=\rho_{A_k}$ for $k=\overline{1,m}$ and any $n$. Thus, it follows from (\ref{delta-I})
that
$$
E_{sq}(\rho_{A_1...A_m})-f_n(\rho_{A_1...A_m})\leq \sum_{k=1}^m \left(2[H(\rho_{A_k})-H(P_k^n\rho_{A_k}P_k^n)]+h_2(\Tr P_k^n\rho_{A_k})\right).
$$
Since $H(\rho_{A_k})<+\infty$ for all $k$, the r.h.s. of the last inequality tends to zero as $n\to+\infty$.

\smallskip

B) Introduce the monotone sequence of functions
\begin{equation}\label{mse-def-star}
  E^d_{sq}(\varrho_{A_1...A_m})=\textstyle\frac{1}{2}\displaystyle\inf_{\varrho_{A_1...A_mE}}I(A_1\!:...:\!A_n|E)_{\varrho},\quad \dim\H_E\leq d,
\end{equation}
where the infimum is over all extensions $\varrho_{A_1...A_mE}$ of the state
$\varrho_{A_1...A_m}$ such that $E$ is a $d$-dimensional system,
pointwise converging as $d\to+\infty$ to the function
\begin{equation}\label{mse-def-star}
  E^*_{sq}(\varrho_{A_1...A_m})=\textstyle\frac{1}{2}\displaystyle\inf_{\varrho_{A_1...A_mE}}I(A_1\!:...:\!A_n|E)_{\varrho},\quad \dim\H_E<+\infty,
\end{equation}
where the infimum is over all extensions $\varrho_{A_1...A_mE}$ of the state
$\varrho_{A_1...A_m}$ such that $E$ is a finite-dimensional system. By direct generalization of the
proof of Lemma 16B in \cite{SE} based on the relation
$$
I(A_1\!:...:\!A_m|E)-I(A_1\!:...:\!A_m)=I(A_1..A_m\!:\!E)-\sum_{k=1}^m I(A_k\!:\!E)
$$
it is easy to show that condition (\ref{mI-cont}) implies that
$$
\lim_{n\to+\infty}E^d_{sq}(\rho^n_{A_1...A_m})=E^d_{sq}(\rho^0_{A_1...A_m})
$$
for any $d$, and hence we have
\begin{equation}\label{E-usc}
\limsup_{n\to+\infty}E^*_{sq}(\rho^n_{A_1...A_m})\leq E^*_{sq}(\rho^0_{A_1...A_m}).
\end{equation}

Introduce also the function
\begin{equation}\label{mse-ue}
  \widehat{E}_{sq}(\varrho_{A_1...A_m})=\sup_{P_{A_1},...,P_{A_m}}E_{sq}(P_{A_1}\otimes...\otimes P_{A_m}\cdot\varrho_{A_1...A_m}\cdot P_{A_1}\otimes...\otimes P_{A_m})
\end{equation}
where the supremum is over all finite-rank projectors $P_{A_1}\in\B(\H_{A_1})$,...,$P_{A_m}\in\B(\H_{A_m})$ and it is assumed that
$E_{sq}(\sigma)=c E_{sq}(\sigma/c)$ for any positive operator $\sigma$ in $\T(\H_{A_1..A_m})$ such that $\Tr\sigma=c>0$. Note that
the function $E_{sq}$ in the r.h.s. of (\ref{mse-ue}) is a finite-dimensional squashed entanglement for any set of projectors $\,P_{A_1},...,P_{A_m}$.

According to the notations in \cite{SE} the function $\widehat{E}_{sq}$ can be called the universal extension of the
finite-dimensional multipartite squashed entanglement. By using the arguments from the proof of Proposition 5 in \cite{SE}  it is easy to show lower semicontinuity and convexity of the function  $\widehat{E}_{sq}$ on $\S(\H_{A_1..A_m})$ and its monotonicity under selective unilocal operations.

It follows from the definitions that
\begin{equation}\label{2-ineq}
E^*_{sq}(\varrho_{A_1..A_m})\geq E_{sq}(\varrho_{A_1..A_m})\geq \widehat{E}_{sq}(\varrho_{A_1..A_m})\quad \forall \varrho_{A_1..A_m}\in\S(\H_{A_1..A_m}).
\end{equation}
We will show that
\begin{equation}\label{2-eq}
E^*_{sq}(\varrho_{A_1..A_m})=E_{sq}(\varrho_{A_1..A_m})=\widehat{E}_{sq}(\varrho_{A_1..A_m})\quad \forall \varrho_{A_1..A_m}\in\S_*(\H_{A_1..A_m}).
\end{equation}
Take arbitrary $\varepsilon>0$ and an extension $\varrho^{\shs\varepsilon}_{A_1..A_mE}$ such that
\begin{equation}\label{2-eq+}
E_{sq}(\varrho_{A_1...A_m})\geq\textstyle\frac{1}{2}\displaystyle I(A_1\!:...:\!A_m|E)_{\varrho^{\shs\varepsilon}}-\varepsilon.
\end{equation}
Let $\{\Phi^n_E\}$ be a sequence of channels from the system $E$ to itself strongly converging to the identity channel $\id_E$ such that
the range of $\Phi^n_E$ is finite-dimensional for each $n$. Since $H(\varrho^{\shs\varepsilon}_{A_k})=H(\varrho_{A_k})<+\infty$ for all $k$,  Proposition 5D in \cite{CMI} shows that
$$
\lim_{n\to+\infty}I(A_1\!:...:\!A_m|E)_{\id_{A_1..A_m}\otimes\Phi^n_E(\varrho^{\shs\varepsilon})}=I(A_1\!:...:\!A_m|E)_{\varrho^{\shs\varepsilon}}.
$$
Since  $\,\frac{1}{2}I(A_1\!:...:\!A_m|E)_{\id_{A_1..A_m}\otimes\Phi^n_E(\varrho^{\varepsilon})}\geq E^*_{sq}(\varrho_{A_1..A_m})\,$ for all $n$ by definition of $E^*_{sq}$, this limit relation and (\ref{2-eq+}) imply the first equality in (\ref{2-eq}).\smallskip

Let $\{P_1^n\}\subset \B(\H_{A_1})$,..., $\{P_m^n\}\subset \B(\H_{A_m})$ be sequences of finite rank projectors strongly converging
to the unit operators $I_{A_1}$,..., $I_{A_m}$. For each sufficiently large $n$ let $\varrho^n_{A_1..A_m}$ be the state proportional to the operator $P_1^n\otimes...\otimes P_m^n\cdot\varrho_{A_1..A_m}\cdot P_1^n\otimes...\otimes P_m^n$. Then the sequence $\{\varrho^n_{A_1..A_m}\}$ belongs to the set
$\S_*(\H_{A_1..A_m})$ and converges to the state $\varrho_{A_1..A_m}\in\S_*(\H_{A_1..A_m})$. By part A of this proposition the function $E_{sq}$ is lower semicontinuous on $\S_*(\H_{A_1..A_m})$. This property and the nonincreasing of $E_{sq}$ under unilocal operations imply that
$$
\lim_{n\to+\infty}E_{sq}(\varrho^n_{A_1..A_m})=E_{sq}(\varrho_{A_1..A_m}).
$$
The same limit relation holds for the function $\hat{E}_{sq}$, since it is lower semicontinuous on the whole space $\S(\H_{A_1..A_m})$ and
nonincreasing under unilocal operations. This implies the second  equality in (\ref{2-eq}), since $E_{sq}(\varrho^n_{A_1..A_m})=\hat{E}_{sq}(\varrho^n_{A_1..A_m})$ for all $n$ (by the construction of $\hat{E}_{sq}$).

Now we are able to prove the first assertion of part B. Indeed, it follows from (\ref{2-ineq}) and (\ref{2-eq}) that
$$
E^*_{sq}(\rho^n_{A_1..A_m})\geq E_{sq}(\rho^n_{A_1..A_m})\geq \widehat{E}_{sq}(\rho^n_{A_1..A_m})
$$
for all $n$ and
$$
E^*_{sq}(\rho^0_{A_1..A_m})=E_{sq}(\rho^0_{A_1..A_m})=\widehat{E}_{sq}(\rho^0_{A_1..A_m}).
$$
Thus, the relation (\ref{E-usc}) and the lower semicontinuity of $\hat{E}_{sq}$ imply (\ref{mSE-cont}).\smallskip

The second assertion of part B follows from the first one and Proposition \ref{main-r+}. $\square$ \smallskip

\begin{lemma}\label{imp-l}
\emph{Let $\rho_{A_1..A_mE}$ be a state such that $H(\rho_{A_k})<+\infty$ for some $k$. Then
$$
|I(A_1\!:...:\!A_m|E)_{\rho}-I(A_1\!:...:\!A_m|E)_{\Phi^n_k\otimes\id_{A^{\rm c}_k\!E}(\rho)} |\leq 2[H(\rho_{A_k})-H(P_k^n\rho_{A_k}P_k^n)]+h_2(\Tr P_k^n\rho_{A_k}),
$$
where $\Phi^n_k:A_k\rightarrow A_k$ is the channel defined in (\ref{phi-k}) and $A^{\rm c}_k=A_1..A_m\setminus A_k$.}
\end{lemma}\smallskip

\emph{Proof.} By using representation (\ref{cmi-mpd+}) with appropriate permutation of the subsystems we obtain
$$
|I(A_1\!:...:\!A_m|E)_{\rho}-I(A_1\!:...:\!A_m|E)_{\Phi^n_k\otimes\id_{A^{\rm c}_k\!E}(\rho)} |=
|I(A_k\!:\!A_k^{\rm c}|E)_{\rho}-I(A_k\!:\!A_k^{\rm c}|E)_{\Phi^n_k\otimes\id_{A^{\rm c}_k\!E}(\rho)}|
$$
Since $\,\Phi^n_k\otimes\id_{A^{\rm c}_kE}(\rho)=P^n_k\otimes I_{A^{\rm c}_kE}\cdot\rho\cdot P^n_k\otimes I_{A^{\rm c}_kE}+\tau^n_k\otimes [\Tr_{A_k}P^n_k\otimes I_{A^{\rm c}_kE}\rho\shs]\,$, inequality (\ref{F-c-b+}) implies that
$$
I(A_k\!:\!A_k^{\rm c}|E)_{\Phi^n_k\otimes\id_{A^{\rm c}_k\!E}(\rho)}\geq I(A_k\!:\!A_k^{\rm c}|E)_{P^n_k\otimes I_{A^{\rm c}_k\!E}\cdot\rho\cdot P^n_k\otimes I_{A^{\rm c}_kE}}-h_2(\Tr P^n_k\rho_{A_k}),
$$
where the vanishing of $I(A_k\!:\!A_k^{\rm c}|E)$ at the state proportional to $\tau^n_k\otimes [\Tr_{A_k}P^n_k\otimes I_{A^{\rm c}_kE}\rho\shs]$ was used.\footnote{We assume that $I(A_k\!:\!A_k^{\rm c}|E)_{\sigma}=cI(A_k\!:\!A_k^{\rm c}|E)_{\sigma\!/\!c}$ for a positive operator $\sigma$ such that $c=\Tr\sigma>0$.}
Thus, we have
$$
\begin{array}{c}
|I(A_1\!:...:\!A_m|E)_{\rho}-I(A_1\!:...:\!A_m|E)_{\Phi^n_k\otimes\id_{A^{\rm c}_k\!E}(\rho)} |\leq
I(A_k\!:\!A_k^{\rm c}|E)_{\rho}\\\\-I(A_k\!:\!A_k^{\rm c}|E)_{P^n_k\otimes I_{A^{\rm c}_k\!E}\cdot\rho\cdot P^n_k\otimes I_{A^{\rm c}_k\!E}}+h_2(\Tr P^n_k\rho_{A_k})\\\\\leq 2[H(\rho_{A_k})-H(P_k^n\rho_{A_k}P_k^n)]+h_2(\Tr P_k^n\rho_{A_k}),
\end{array}
$$
where the last inequality follows from Lemma 9 in \cite{SE}. $\square$\smallskip

\begin{corollary}\label{mSE-0} \emph{The function $E_{sq}$ defined in (\ref{mse-def}) is equal to zero on the set of all separable states $\rho_{A_1...A_m}$
with finite marginal entropies, i.e. such that $H(\rho_{A_k})<+\infty$ for $k=\overline{1,m}$.}
\end{corollary}\smallskip

\emph{Proof.} Let $\{P_1^n\}\subset \B(\H_{A_1})$,..., $\{P_m^n\}\subset \B(\H_{A_m})$ be sequences of finite rank projectors strongly converging
to the unit operators $I_{A_1}$,..., $I_{A_m}$. For each sufficiently large $n$ let $\rho^n_{A_1..A_m}$ be the state proportional to the operator $P_1^n\otimes...\otimes P_m^n\cdot\rho_{A_1..A_m}\cdot P_1^n\otimes...\otimes P_m^n$. Since $\rho^n_{A_1..A_m}$ is a finite-dimensional separable state,
$E_{sq}(\rho^n_{A_1..A_m})=0$ for all $n$. Since the function $E_{sq}$ is lower semicontinuous on the set $\S_*(\H_{A_1..A_m})$ defined in (\ref{Star}) by Proposition \ref{SE-m}A,  we have $E_{sq}(\rho_{A_1..A_m})=0$. $\square$ \smallskip

Since the function $E_{sq}$ is convex and equal to zero on the set of countably-decomposable separable states, i.e. states $\sigma_{A_1...A_m}$ having the form
$$
\sigma_{A_1...A_m}=\sum_i p_i\shs \alpha^i_1\otimes...\otimes\alpha^i_m,
$$
the assertion of Corollary \ref{mSE-0} can be strengthened by saying that $E_{sq}(\rho_{A_1...A_m})=0$ if $\rho_{A_1...A_m}$ is
a mixture of a countably-decomposable separable state and a separable state with finite marginal entropies.
\medskip

Corollary \ref{main-r+c} and Proposition \ref{SE-m} imply
the following condition for local continuity of
the multipartite squashed entanglement.\smallskip

\begin{corollary}\label{SE-cont-cond} \emph{Let $\{\rho^n_{A_1...A_m}\}$ be a sequence of states converging to a state
$\rho^0_{A_1...A_m}$ such that $H(\rho^0_{A_k})$ is finite for $k=\overline{1,m}$. If there exist systems $E_1,...E_{m-1}$ and a sequence $\{\hat{\rho}_{A_1...A_mE_1...E_{m-1}}^n\}$ of states converging to a state $\hat{\rho}_{A_1...A_mE_1...E_{m-1}}^0$ such that
$\hat{\rho}_{A_1...A_m}^n=\rho_{A_1...A_m}^n$ for all $\,n\geq 0$ and $\,\lim_{n\rightarrow+\infty}H(\rho^n_{A_{i_k}E_k})=H(\rho^0_{A_{i_k}E_k})<+\infty$ for some set of indexes $i_1,...,i_{m-1}$ then relation  (\ref{mSE-cont}) holds.}
\end{corollary}\smallskip

By generalizing Example \ref{e-2} consider arbitrary sequences $\{\rho^n_{A_1}\}$,...,$\{\rho^n_{A_m}\}$ and $\{\sigma^n_{A_1...A_m}\}$ of states converging, respectively, to states $\{\rho^0_{A_1}\}$,...,$\{\rho^0_{A_m}\}$ and $\{\sigma^0_{A_1...A_m}\}$. By obvious generalization of the 
arguments from Example \ref{e-2} one can apply Corollary \ref{SE-cont-cond}  to show that
$$
\lim_{n\rightarrow+\infty} E_{sq}((1-p)\rho^n_{A_1}\otimes...\otimes\rho^n_{A_m}+p\,\sigma^n_{A_1...A_m})=E_{sq}((1-p)\rho^0_{A_1}\otimes...\otimes\rho^0_{A_m}
+p\,\sigma^0_{A_1...A_m})
$$
for any $p\in[0,1]$ provided that
\begin{equation}\label{s-cond+}
\lim_{n\rightarrow+\infty}H(\sigma^n_{A_k})=H(\sigma^0_{A_k})<+\infty\quad \textrm{and}\quad H(\rho^0_{A_k})<+\infty\quad \textrm{for}\;\, k=\overline{1,m}
\end{equation}
\emph{regardless} of the entropies of the states $\rho^n_{A_1}$,...,$\rho^n_{A_m}$.

The second condition in (\ref{s-cond+}) seems technical (i.e. it can be omitted by using more advanced 
methods of analysis) but the first condition in (\ref{s-cond+}) is essential. This can be shown by     
considering the case when $p=1$ and all the states $\sigma^n_{A_1...A_m}$ are pure.

\subsection{On robustness of information gain of quantum measurements}

A quantum measurement with finite or countable outcome set $I$ is described mathematically
by a quantum instrument $\mathbf{\Upsilon}=\{\Upsilon_i\}_{i\in I}$ -- collection of quantum operations from a system $A$ to a system $A'$ such that
$\sum_{i\in I}\Tr\Upsilon_i(\rho)=1$ for any state $\rho$ in $\S(\H_A)$. Many characterictics of a quantum measurement can be expressed
via the corresponding Positive Operator Valued Measure (POVM) $\{\Upsilon^*_i(I_{\H_{A'}})\}_{i\in I}$.
A quantum measurement is called efficient if all the quantum operations $\Upsilon_i$ have Choi rank 1 \cite{H-SSQT,N&Ch,Wilde}.

Following to Groenewold \cite{G} the information gain of an efficient measurement at a state $\rho$ is defined as
\begin{equation}\label{ER}
  H(\rho)-\sum_{i\in I}p_iH(\rho_i),
\end{equation}
where $p_i=\Tr \Upsilon_i(\rho)$ is the probability of the $i$-th outcome and  $\rho_i=p_i^{-1}\Upsilon_i(\rho)$ is the corresponding
posteriory state. This quantity naturally called the entropy reduction  is nonnegative for
any efficient measurement \cite{L-3,Ozawa}.

For an arbitrary non-efficient measurement the entropy reduction (\ref{ER}) may be negative and can not be used as its information characteristic. In this case the
infromation gain of a quantum measurement described by the instrument $\{\Upsilon_i\}_{i\in I}$ at a state $\rho$ in $\S(\H_A)$ is given by the quantum mutual information $I(R\!:\!E)$
of the state
$$
\sigma_{RE}=\sum_{i\in I}\Tr_{A'} [\shs\id_R\otimes \Upsilon_i(\hat{\rho}_{AR})]\otimes |i_E\rangle\langle i_E|,
$$
where $\hat{\rho}_{AR}$ is a purification of the state $\rho$ and $\{|i_E\rangle\}$ is a basis in a Hilbert space $\H_E$ such that $\dim\H_E=\mathrm{card}(I)$ \cite{GIB}.
It is easy to see that for any efficient measurement this quantity coincides with the entropy reduction (\ref{ER}).

Winter showed that the above quantity $I(R\!:\!E)_{\sigma}$ has an operational interpretation as the
optimal rate at which a measurement gathers information \cite{W-IG}. This result has been generalized in \cite{IGSI}
to the case when a measurement is applied to part A of an entangled bipartite state $\rho_{AB}$. It is shown that
in this case the optimal rate at which the sender needs to transmit classical information in order to simulate the measurement is equal to
the quantum conditional mutual information $I(R\!:\!E|B)_{\sigma}$ of the state
$$
\sigma_{REB}=\sum_{i\in I}\Tr_{A'} [\shs \id_{RB}\otimes \Upsilon_i(\hat{\rho}_{AR})]\otimes |i_E\rangle\langle i_E|,
$$
where $\hat{\rho}_{ARB}$ is a purification of the state $\rho_{AB}$. This quantity is called the \emph{information gain in the presence of quantum side information}, while the above quantity $I(R\!:\!E)_{\sigma}$ -- the \emph{information gain without quantum side information} \cite{IGSI+,ED-2,IGSI}.
\smallskip

In this subsection we  analyse continuity properties of the characteristics of quantum
measurements described above. Note first that both these characteristics are completely determined by
the POVM $\{\Upsilon^*_i(I_{\H_{A'}})\}_{i\in I}$ corresponding to the instrument $\{\Upsilon_i\}_{i\in I}$.
Indeed, for a given POVM $\M=\{M_i\}_{i\in I}$ consider the channel
\begin{equation}\label{psi-ch}
\Psi_{\M}(\rho)=\sum_{i}[\Tr M_i\rho]|i_E\rangle\langle i_E|
\end{equation}
from $\S(\H_{A})$ to $\S(\H_{E})$. Then the above defined states  $\sigma_{RE}$ and $\sigma_{REB}$ can be expressed as
$$
\sigma_{RE}=\Psi_{\M_{\mathbf{\Upsilon}}}\otimes \id_{R}(\hat{\rho}_{AR})\quad \textrm{and} \quad \sigma_{REB}=\Psi_{\M_{\mathbf{\Upsilon}}}\otimes \id_{BR}(\hat{\rho}_{ABR}),
$$
where $\M_{\mathbf{\Upsilon}}=\{\Upsilon^*_i(I_{\H_{A'}})\}_{i\in I}$.\smallskip

Thus, we may consider the information gain with and without quantum side information as functions of a pair $(\M, \rho_X)$, where $\M$ runs
over all POVM on $\H_A$ with a given set of outcomes and $\rho_X$ runs over the set $\S(\H_X)$, $X=A, AB$. We will use the notations
$$
\mathrm{IG}(\M,\rho_A)=I(R\!:\!E)_{\Psi_{\M}\otimes \id_{R}(\hat{\rho}_{AR})}\quad \textrm{and} \quad\mathrm{IG}_{\rm QSI}(\M,\rho_{AB})=I(R\!:\!E|B)_{\Psi_{\M}\otimes \id_{BR}(\hat{\rho}_{ABR})},
$$
where $\hat{\rho}_{AR}$ and $\hat{\rho}_{ABR}$ are purifications of the states $\rho_A$ and $\rho_{AB}$ correspondingly. \smallskip

To obtain continuity conditions in the strongest form we equip the set $\mathfrak{M}_A$ of all POVM on $\H_A$ with a given outcome set $I$ with the \emph{weak
convergence}. We say that a sequence  $\{\M_n=\{M^n_i\}\}_n$ in $\mathfrak{M}_A$ weakly converges to a POVM $\M_0=\{M^0_i\}\in\mathfrak{M}_A$  if
the sequence  $\{M^n_i\}_n$ converges to the operator $M^0_i$ in the weak operator topology for each $i\in I$. The compactness of the unit ball of
$\B(\H_A)$ in the weak operator topology implies that the set $\mathfrak{M}_A$ is compact if the set $I$ of outcomes is finite. \smallskip

We will essentially use the following\smallskip

\begin{lemma}\label{vsl} \emph{If  a sequence  $\{\M_n\}$ in $\mathfrak{M}_A$ weakly converges to a POVM $\M_0\in\mathfrak{M}_A$ then
the sequence $\{\Psi_{\M_n}\}$ of channels defined in (\ref{psi-ch}) strongly converges to the channel $\Psi_{\M_0}$, i.e. }
$$
\lim_{n\rightarrow+\infty}\Psi_{\M_n}(\rho)=\Psi_{\M_0}(\rho)\quad \forall \rho\in\S(\H_A).
$$
\end{lemma}

\emph{Proof.} By Lemma \ref{D-A} it suffices to show that the sequence
$\{\Psi_{\M_n}(\rho)\}$ converges to the state $\Psi_{\M_0}(\rho)$ in the weak operator topology for any given  state
$\rho$. This can be done easily by noting that the weak convergence of the sequence $\{\M_n\}$ to a POVM $\M_0$ implies that
$\lim_{n\rightarrow+\infty}\Tr M^n_i\rho=\Tr M^0_i\rho$ for each $i\in I$. $\square$\smallskip

Now we can prove the following\smallskip

\begin{property}\label{IGR} A) \emph{Let be $\{\rho^n_A\}$ be a sequence of states in $\S(\H_{A})$ converging to a state $\rho_A^0$. If either
$\,\mathrm{card}(I)<+\infty\,$ or $\,\lim_{n\rightarrow+\infty}H(\rho^n_A)=H(\rho^0_A)<+\infty$ then
$$
\lim_{n\rightarrow+\infty}\mathrm{IG}(\M_n,\rho_{A}^n)=\mathrm{IG}(\M_0,\rho_{A}^0)
$$
for arbitrary sequence $\{\M_n\}$ in $\mathfrak{M}_A$ weakly converging to a POVM $\M_0\in\mathfrak{M}_A$.}\smallskip

B) \emph{Let $\{\rho^n_{AB}\}$ be a sequence of states in $\S(\H_{AB})$ converging to a state $\rho_{AB}^0$. If either
$\,\mathrm{card}(I)<+\infty\,$ or $\,\lim_{n\rightarrow+\infty}H(\rho^n_{X})=H(\rho^0_{X})<+\infty$, where $X$ is either $A$ or $AB$, then
$$
\lim_{n\rightarrow+\infty}\mathrm{IG}_{\rm QSI}(\M_n,\rho_{AB}^n)=\mathrm{IG}_{\rm QSI}(\M_0,\rho_{AB}^0)
$$
for arbitrary sequence $\{\M_n\}$ in $\mathfrak{M}_A$ weakly converging to a POVM $\M_0\in\mathfrak{M}_A$.}
\end{property}\smallskip

\emph{Proof.} Since  $\dim\H_E=\mathrm{card}(I)$, all the assertions of the proposition follow, by Lemma \ref{vsl}, from Corollary \ref{cmi-th-c-1} and Proposition \ref{main-r}. $\square$\smallskip

Proposition \ref{IGR} implies, in particular, that for \emph{arbitrary} sequence $\{\M_n\}$ weakly converging to a POVM $\M_0$
we have
$$
\lim_{n\rightarrow+\infty}\mathrm{IG}(\M_n,\rho_{A})=\mathrm{IG}(\M_0,\rho_{A})
$$
provided that $H(\rho_{A})<+\infty$ and that
$$
\lim_{n\rightarrow+\infty}\mathrm{IG}_{\rm QSI}(\M_n,\rho_{AB})=\mathrm{IG}_{\rm QSI}(\M_0,\rho_{AB})
$$
provided that either $H(\rho_{AB})<+\infty$ or $H(\rho_{A})<+\infty$. This property can be treated as
stability (robustness) the information gain with and without quantum side information w.r.t. perturbation of quantum
measurements.\smallskip

A sequence of quantum instruments $\mathbf{\Upsilon}_n=\{\Upsilon_i^n\}$ with the same outcomes set $I$  \emph{strongly converges} to
a quantum instrument $\mathbf{\Upsilon}_0=\{\Upsilon^0_i\}$ if
$$
\lim_{n\rightarrow+\infty}\Upsilon^n_i(\rho)=\Upsilon^0_i(\rho)
$$
for each $i\in I$ and any state $\rho$ in $\S(\H_A)$ \cite{ER}. It is clear that this convergence implies
the weak convergence of the corresponding sequence of POVM $\M_{\mathbf{\Upsilon}_n}=\{[\Upsilon_i^n]^*(I_{A'})\}$ to the
POVM $\M_{\mathbf{\Upsilon}_0}=\{[\Upsilon_i^0]^*(I_{A'})\}$. Hence, Proposition \ref{IGR} implies the following  \smallskip

\begin{corollary}\label{IGR-c} A) \emph{Let $\{\rho^n_A\}$ be a sequence of states in $\,\S(\H_{A})$ converging to a state $\rho_A^0$.
If either
$\,\mathrm{card}(I)<+\infty\,$ or $\,\lim_{n\rightarrow+\infty}H(\rho^n_A)=H(\rho^0_A)<+\infty$ then
$$
\lim_{n\rightarrow+\infty}\mathrm{IG}(\M_{\mathbf{\Upsilon}_n},\rho_{A}^n)=\mathrm{IG}(\M_{\mathbf{\Upsilon}_0},\rho_{A}^0)
$$
for arbitrary sequence $\{\mathbf{\Upsilon}_n\}$ of quantum instruments strongly converging to a quantum instrument $\mathbf{\Upsilon}_0$.}\smallskip

B) \emph{Let $\{\rho^n_{AB}\}$ be a sequence of states in $\,\S(\H_{AB})$ converging to a state $\rho_{AB}^0$. If either
$\,\mathrm{card}(I)<+\infty\,$ or $\,\lim_{n\rightarrow+\infty}H(\rho^n_{X})=H(\rho^0_{X})<+\infty$, where $X$ is either $A$ or $AB$, then
$$
\lim_{n\rightarrow+\infty}\mathrm{IG}_{\mathrm{QSI}}(\M_{\mathbf{\Upsilon}_n},\rho_{AB}^n)=\mathrm{IG}_{\mathrm{QSI}}(\M_{\mathbf{\Upsilon}_0},\rho_{AB}^0)
$$
for arbitrary sequence $\{\mathbf{\Upsilon}_n\}$ of quantum instruments strongly converging to a quantum instrument $\mathbf{\Upsilon}_0$.}
\end{corollary}\smallskip

Corollary \ref{IGR-c}A agrees with Proposition 1 in \cite{ER}, since if an  instrument $\mathbf{\Upsilon}$  is efficient then
the information gain $\,\mathrm{IG}(\M_{\mathbf{\Upsilon}},\rho_{A})$ coincides with the entropy reduction (\ref{ER}).

\bigskip

I am grateful to A.S.Holevo and to the participants of his seminar
"Quantum probability, statistic, information" (the Steklov
Mathematical Institute) for useful discussion.

\end{document}